\documentclass{aa}

\usepackage{graphicx}
\usepackage{lipsum}

\usepackage{txfonts}
\usepackage{natbib}
\usepackage{hyperref}
\usepackage{longtable}
\usepackage{supertabular}
\usepackage{placeins}

\usepackage[normalem]{ulem}

\usepackage{caption}
\usepackage{subcaption}

\usepackage{lscape}

\usepackage[utf8]{inputenc}
\usepackage{booktabs}
\usepackage{hyperref}
\usepackage{multirow}
\usepackage{enumitem}

\usepackage{lscape}

\usepackage[switch]{lineno}

\usepackage[dvipsnames]{xcolor}
\definecolor{myblue}{HTML}{1F77B4}
\definecolor{mygreen}{HTML}{2CA02C}
\definecolor{m}{HTML}{D62728}
\definecolor{mymagenta}{HTML}{D33682}
\definecolor{codepurple}{HTML}{C42043}

\newcommand{\program}[1]{\textsc{#1}}

\hypersetup{
unicode=true,           % non-Latin %characters in Acrobat’s bookmarks
colorlinks=true,        % false: boxed %links; true: colored links
citecolor=myblue,       % color of links %to bibliography
urlcolor=black        % color of %external links
}
\begin{document}

   \title{SN\,2020zbf: A fast-rising hydrogen-poor superluminous supernova with strong carbon lines}

   \author{A. Gkini\inst{1} \href{https://orcid.org/0009-0000-9383-2305}{\includegraphics[scale=0.5]{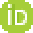}}\and 
    R. Lunnan\inst{1} \href{https://orcid.org/0000-0001-9454-4639}{\includegraphics[scale=0.5]{Images/ORCIDiD_icon16x16.eps}}\and 
    S. Schulze\inst{2} \href{https://orcid.org/0000-0001-6797-1889}{\includegraphics[scale=0.5]{Images/ORCIDiD_icon16x16.eps}} \and
    L. Dessart \inst{3} \href{https://orcid.org/0000-0003-0599-8407}{\includegraphics[scale=0.5]{Images/ORCIDiD_icon16x16.eps}}\and 
    S. J. Brennan\inst{1} \href{https://orcid.org/0000-0003-1325-6235}{\includegraphics[scale=0.5]{Images/ORCIDiD_icon16x16.eps}}
    \and 
    J. Sollerman\inst{1} \href{https://orcid.org/0000-0003-1546-6615}{\includegraphics[scale=0.5]{Images/ORCIDiD_icon16x16.eps}} \and
    P. J. Pessi\inst{1} \href{https://orcid.org/0000-0002-8041-8559}{\includegraphics[scale=0.5]{Images/ORCIDiD_icon16x16.eps}} \and 
    M. Nicholl\inst{4} \href{https://orcid.org/0000-0002-2555-3192}{\includegraphics[scale=0.5]{Images/ORCIDiD_icon16x16.eps}} \and
    L. Yan \inst{5} \and
    C. M. B. Omand\inst{1} \href{https://orcid.org/0000-0002-9646-8710}{\includegraphics[scale=0.5]{Images/ORCIDiD_icon16x16.eps}} \and
     T. Kangas\inst{6,7} \href{https://orcid.org/0000-0002-5477-0217}{\includegraphics[scale=0.5]{Images/ORCIDiD_icon16x16.eps}}\and 
    T. Moore \inst{4} \href{https://orcid.org/0000-0001-8385-3727}{\includegraphics[scale=0.5]{Images/ORCIDiD_icon16x16.eps}}\and
    J.~P. Anderson \inst{8,9} \href{https://orcid.org/0000-0003-0227-3451}{\includegraphics[scale=0.5]{Images/ORCIDiD_icon16x16.eps}}\and
     T.-W. Chen\inst{10} \href{https://orcid.org/0000-0002-1066-6098}{\includegraphics[scale=0.5]{Images/ORCIDiD_icon16x16.eps}} \and
     E. P. Gonzalez \inst{11,12} \and
     M. Gromadzki \inst{13} \href{https://orcid.org/0000-0002-1650-1518}{\includegraphics[scale=0.5]{Images/ORCIDiD_icon16x16.eps}} \and
     C. P. Guti\'errez \inst{14,15} \and
     D. Hiramatsu \inst{16,17} \href{https://orcid.org/0000-0002-1125-9187}{\includegraphics[scale=0.5]{Images/ORCIDiD_icon16x16.eps}}\and
    D. A. Howell   \inst{11,12}  \href{https://orcid.org/0000-0003-4253-656X}{\includegraphics[scale=0.5]{Images/ORCIDiD_icon16x16.eps}} 
    \and
    N. Ihanec  \inst{13} 
     \and
     C. Inserra \inst{18} \href{https://orcid.org/0000-0002-3968-4409}{\includegraphics[scale=0.5]{Images/ORCIDiD_icon16x16.eps}} \and
    C. McCully   \inst{11} \href{https://orcid.org/0000-0001-5807-7893}{\includegraphics[scale=0.5]{Images/ORCIDiD_icon16x16.eps}}  \and
     T. E. M\"uller-Bravo \inst{15,14} \href{https://orcid.org/0000-0003-3939-7167}{\includegraphics[scale=0.5]{Images/ORCIDiD_icon16x16.eps}} \and
    C. Pellegrino \inst{11,12} \href{https://orcid.org/0000-0002-7472-1279}{\includegraphics[scale=0.5]{Images/ORCIDiD_icon16x16.eps}} \and
     G. Pignata \inst{19} \href{https://orcid.org/0000-0003-0006-0188}{\includegraphics[scale=0.5]{Images/ORCIDiD_icon16x16.eps}} \and
     M. Pursiainen\inst{20} \href{https://orcid.org/0000-0002-8597-0756}{\includegraphics[scale=0.5]{Images/ORCIDiD_icon16x16.eps}} \and
     D. R. Young\inst{4} \href{https://orcid.org/0000–0002–1229–2499}{\includegraphics[scale=0.5]{Images/ORCIDiD_icon16x16.eps}}}

   \institute{The Oskar Klein Centre, Department of Astronomy, Stockholm University, Albanova University Center, 106 91 Stockholm, Sweden
         \and
             The Oskar Klein Centre, Department of Physics, Stockholm University, Albanova University Center, 106 91 Stockholm, Sweden 
    \and
    Institut d’Astrophysique de Paris, CNRS-Sorbonne Université, 98 bis boulevardArago, 75014 Paris, France %3
    \and
    Astrophysics Research Centre, School of Mathematics and Physics, Queens University Belfast, Belfast BT7 1NN, UK %4
    \and
    The Caltech Optical Observatories, California Institute of Technology, Pasadena, CA 91125, USA %5
    \and
    Finnish Centre for Astronomy with ESO (FINCA), University of Turku, 20014 Turku, Finland %6
    \and
    Tuorla Observatory, Department of Physics and Astronomy, University of Turku, 20014 Turku, Finland %7
    \and
    European Southern Observatory, Alonso de C\'ordova 3107, Casilla 19, Santiago, Chile %8
    \and
    Millennium Institute of Astrophysics MAS, Nuncio Monsenor Sotero Sanz 100, Of. 104, Providencia, Santiago, Chile %9
    \and
    Graduate Institute of Astronomy, National Central University, 300 Jhongda Road, 32001 Jhongli, Taiwan %10
     \and
    Las Cumbres Observatory, 6740 Cortona Dr. Suite 102, Goleta, CA, 93117, USA %11
    \and
    Department of Physics, University of California, Santa Barbara, CA 93106-9530, USA %12
    \and
    Astronomical Observatory, University of Warsaw, Al. Ujazdowskie 4, PL-00-478 Warszawa, Poland %13
    \and
     Institut d’Estudis Espacials de Catalunya (IEEC), Gran Capit\`a, 2-4,
    Edifici Nexus, Desp. 201, E-08034 Barcelona, Spain%14
    \and
    Institute of Space Sciences (ICE, CSIC), Campus UAB, Carrer de Can
    Magrans, s/n, E-08193 Barcelona, Spain
     %15
    \and
    Center for Astrophysics, Harvard \& Smithsonian, 60 Garden Street, Cambridge, MA 02138-1516, USA %16
    \and
    The NSF AI Institute for Artificial Intelligence and Fundamental Interactions, USA %17
     \and
    Cardiff Hub for Astrophysics Research and Technology, School of Physics \& Astronomy, Cardiff University, Queens Buildings, The Parade, Cardiff, CF24 3AA, UK %18
    \and
     Instituto de Alta Investigación, Universidad de Tarapacá, Casilla 7D, Arica, Chile %19
     \and
   Department of Physics, University of Warwick, Gibbet Hill Road, Coventry CV4 7AL, UK %20
   }
   \date{}

\abstract
{SN\,2020zbf is a hydrogen-poor superluminous supernova (SLSN) at $z =  0.1947$ that shows conspicuous \ion{C}{II} features at early times, in contrast to the majority of H-poor SLSNe. Its peak magnitude is $M_{\rm g}$ = $-21.2$~mag  and its rise time ($\lesssim 26.4$ days from first light) places SN\,2020zbf among the fastest rising type I SLSNe. We used spectra taken from ultraviolet (UV) to near-infrared wavelengths to identify spectral features. We paid particular attention to the \ion{C}{II} lines as they present distinctive characteristics when compared to other events. We also analyzed UV and optical photometric data and modeled the light curves considering three different powering mechanisms: radioactive decay of $^{56}$Ni, magnetar spin-down, and circumstellar medium (CSM) interaction. The spectra of SN\,2020zbf match the model spectra of a C-rich low-mass magnetar-powered supernova model well. This is consistent with our light curve modeling, which supports a magnetar-powered event with an ejecta mass $M_{\rm ej}$ = 1.5~$\rm M_\odot$. However, we cannot discard the CSM-interaction model as it may also reproduce the observed features. The interaction with H-poor, carbon-oxygen CSM near peak light could explain the presence of \ion{C}{II} emission lines. A short plateau in the light curve around 35 -- 45 days after peak, in combination with the presence of an emission line at 6580~\AA,\ can also be interpreted as being due to a late interaction with an extended H-rich CSM. Both the magnetar and CSM-interaction models of SN\,2020zbf indicate that the progenitor mass at the time of explosion is between 2 and 5~$\rm M_\odot$.  Modeling the spectral energy distribution of the host galaxy reveals a host mass of 10$^{8.7}$~$\rm M_\odot$, a star formation rate of 0.24$^{+0.41}_{-0.12}$~$\rm M_\odot$~yr$^{-1}$, and a metallicity of $\sim$ 0.4~$\rm Z_\odot$.}


\authorrunning{Gkini, et al.}
\titlerunning{}

   \maketitle
%
%-------------------------------------------------------------------
\section{Introduction}

Modern time-domain sky surveys with large fields of view are able to detect and follow rare transient events. 
Superluminous supernovae (SLSNe) are an extremely luminous class of transients, 10 -- 100 times brighter than canonical supernova (SN) explosions \citep{Quimby2011,GalYam2012}. The need for a new class of SN arose due to the fact that some events \citep{Nugent1999,Ofek2007,Smith2007,Quimby2007,Miller2009,Barbary2009,GalYam2009,Gezari2009} are much brighter than the majority of previously discovered events and could not fit into the conventional SN explosion scenario. SLSNe are frequently detected in metal-poor dwarf host galaxies \citep{Neill2011,Chen2013,Lunnan2014,Leloudas2015,Angus2016,Perley2016,Chen2017b,Schulze2018} and were originally defined as having an absolute magnitude of M $< -21$ \citep{GalYam2012}. However, many SLSNe have peak magnitudes less than this threshold \citep{Inserra2013,Angus2019,Lunnan2018,DeCia2018,Inserra2018a,Chen2023a}, and SLSN classification is better determined using spectroscopic properties \citep{Quimby2011,Quimby2018,GalYam2019a} rather than an arbitrary brightness cut.

The SLSN class is typically divided into two subgroups based on the presence of hydrogen in their spectra around the peak -- type I (H-poor; SLSN-I hereafter) and type II (H-rich; SLSN-II) -- although a few SLSNe-I  have H$\alpha$ emission detected in their late-time spectra \citep{Yan2015,Yan2017,Chen2018,Fiore2021,Pursiainen2022}. In particular, SLSNe-I whose spectra do not show \ion{He}{} are characterized as SLSNe-Ic. It has been proposed that SLSNe-I can be further classified photometrically: ``slow-evolving'' SLSNe-I show a rise of about 50 days and a decline consistent with the $^{56}$\ion{Co}{} decay rate, whereas ``fast-evolving'' SLSNe-I have rise times of less than 30 days \citep{Nicholl2016a,Quimby2018,Inserra2019}. However, with some events being in the intermediate regime (e.g Gaia16apd, \citealt{Kangas2017}; SN\,2017gci, \citealt{Fiore2021}), there are claims of a continuous distribution rather than distinct subclasses \citep{Nicholl2016a,DeCia2018}.

The dominating powering mechanism for SLSNe-II is likely interaction with a dense circumstellar medium (CSM; \citealt{Ofek2014,Inserra2018}, but see \citealt{Kangas2022}). However, for H-poor SLSNe \citep{Chevalier2011,Ofek2013,Inserra2018a}, the powering mechanism is still poorly understood. The amount of radioactive $^{56}$Ni produced in the standard core collapse mechanism is insufficient to explain the extreme brightness of SLSNe-I; thus, other mechanisms have been proposed. One proposed mechanism is pair-instability supernovae (PISNe), in which the formation of positron--electron pairs in the CO core of a $140$ -- $260$~$\rm M_\odot$ zero-age main sequence (ZAMS) metal-poor star results in explosive O burning, and the energy released reverses the collapse and entirely disrupts the star \citep{Woosley2002,Heger2002}. This process has the potential to generate the enormous amount of $^{56}$Ni required to power a SLSN-I light curve. Despite the presence of a few PISN candidates \citep{GalYam2009,Schulze2023}, numerous earlier studies have demonstrated that $^{56}$Ni is not the dominant source of energy for the majority of SLSNe-I \citep{Chatzopoulos2012,Chen2013,Inserra2013,Nicholl2017,Inserra2018,Moriya2018,Chen2023b}. The majority of the observed photometric features of SLSNe-I  \citep{Inserra2013,Nicholl2017,Chen2023b,Omand2024} can instead be attributed to the spin-down of a rapidly rotating young neutron star \citep{Ostriker1971,Kasen2010,Woosley2010}, in which the photons from the newborn magnetar wind nebula are absorbed and thermalized in the SN ejecta, increasing the temperature of the ejecta and the luminosity of the SN.
The spectroscopic signatures of magnetar-powered SLSNe-I have yet to be explored in detail, but simulations demonstrate that magnetars can reproduce the observed SLSN-I spectra \citep{Dessart2012,Mazzali2016,Jerkstrand2017b,Dessart2019,Omand2023a}.

The majority of the early SLSN-I spectra exhibit a steep blue continuum \citep{Yan2017b,Yan2018} and present distinct spectroscopic key characteristics \citep{Quimby2011,Mazzali2016,Quimby2018}. The presence of \ion{O}{II} features dominates the spectra at $3500$ -- $5000$~\AA, with the most significant W-shape feature at $4350$ -- $4650$~\AA,\ which is not typically seen in normal SNe-Ic.  However, numerous SLSNe-I in the literature do not appear to have the W-shape \ion{O}{II} in their spectra \citep[e.g.,][]{Gutirrez2022}, suggesting a further division of the SLSN-I class \citep{Konyves2021}. The red part of the optical SLSN-I spectra presents weak C and O lines, which  \cite{Dessart2012} and \cite{Howell2013} suggest
come from the explosion of the CO core. 
The spectra at $\sim$ 30 days resemble those of SNe-Ic around maximum light \citep{Pastorello2010}.

Interestingly, several SLSNe-I in the literature do not fit into this ``standard'' classification scheme. These events show strong \ion{C}{II} lines in their spectra \citep{Yan2017,Quimby2018,Anderson2018,Fiore2021,Gutirrez2022}. \cite{Anderson2018} suggest that the persistent \ion{C}{II} features in SN\,2018bsz are produced by a magnetar-powered explosion of a massive C-rich Wolf-Rayet (WR) progenitor. The models of \cite{Fiore2021} suggest that the C-rich SN\,2017gci was powered by either a magnetar or CSM interaction with a 40~$\rm M_\odot$ progenitor. Additionally, \cite{Gutirrez2022} find that SN\,2020wnt requires over 4~$\rm M_\odot$ of $^{56}$Ni to produce the observed light curve, which is consistent with the PISN scenario, while \cite{Tinyanont2023} favor a magnetar model. Various ideas have been proposed to explain these characteristics, but the reasons for the presence of the \ion{C}{II} lines are still poorly understood.

In this paper we present an extensive dataset for SN\,2020zbf, a fast-rising SLSN-I with peculiar features in its early spectrum that initially led to an incorrect redshift estimation. A medium-resolution X-shooter spectrum displays three strong \ion{C}{II} lines, indicating that SN\,2020zbf belongs to the C-rich SLSN-I subclass. We performed an extensive investigation of the observed data, modeling the light curve and comparing the high-quality spectra with synthetic models. This allowed us to explore various combinations of power sources and progenitor stars that could result in these spectral signatures. This object, along with other rare SLSNe, may challenge the conventional classification scheme by demonstrating how diverse even the SLSN-I class may be, with implications for both progenitor populations and explosion mechanisms.

This paper is structured as follows. In Sect.~\ref{sec:obs} we present the photometric and spectroscopic observations of SN\,2020zbf as well as the photometric measurements of the host galaxy. In Sect.~\ref{sec:lc_analysis} we analyze the light curve properties, compare them with those of well-studied SLSNe-I, and apply blackbody fits to derive the photospheric temperatures and radius. In Sect.~\ref{sec:spectra} we analyze the spectral properties of SN\,2020zbf. We compare the light curves and the early and the late photospheric spectra with those of typical SLSNe-I in the literature as well as C-rich objects in Sect.~\ref{comparison}. In Sect.~\ref{sec:synthetic_spectra} we compare existing synthetic spectra with our high-quality X-shooter spectrum. We model the multiband light curves of SN\,2020zbf under the assumption that they are powered by three distinct power sources in Sect.~\ref{sec:lc_model}. In Sect.~\ref{sec:host} we discuss the properties of the host galaxy. We discuss the results and provide possible scenarios in Sect.~\ref{sec:discussion} and summarize our findings in Sect.~\ref{sec:conclusions}. Throughout this paper we assume a flat Lambda cold dark matter cosmology with H$_{\rm 0}$ = 67.4~km~s$^{-1}$ Mpc$^{-1}$, $\Omega_{M} = 0.31,$ and $\Omega_{\Lambda} = 0.69 $ \citep{Planck2020}.

%--------------------------------------------------------------------
\section{Observations} \label{sec:obs}

\subsection{Discovery and classification}

\begin{figure*}
    \centering
    \begin{tabular}{cc}
    \includegraphics[width=3.0in]{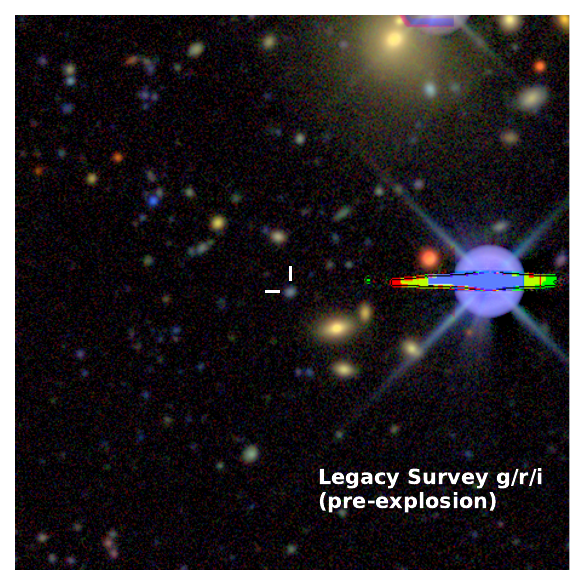} & \includegraphics[width=3.0in]{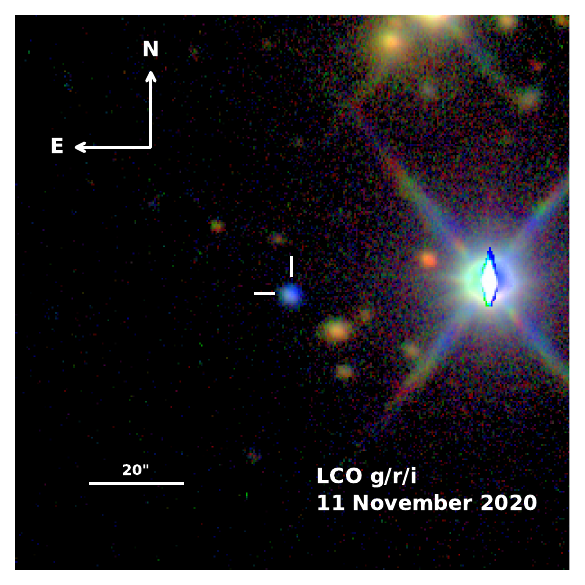}
    \end{tabular}
    \caption{Images of the field of SN\,2020zbf. Left: Legacy Survey DR10 image of the field of SN\,2020zbf before explosion. A faint host galaxy at the SN position is visible, marked by the white crosshairs. Right: $gri$ composite image of the SN near peak from Las Cumbres Observatory (LCO). Both images have a size of $2 \times 2$ arcminutes and have been combined following the algorithm in \citet{Lupton2004}.}
    \label{fig:2020zbf_image}
\end{figure*}

SN\,2020zbf was discovered by the Asteroid Terrestrial-impact Last Alert System (ATLAS; \citealt{Tonry2020}) on November 8, 2020, as ATLAS20bfee at an orange-band magnitude of 18.92~mag at right ascension, declination (J2000) 01$^{h}$58$^{m}$01.65$^{s}$, $-41^{\circ}20'51.89''$. It was classified by the extended Public ESO Spectroscopic Survey for Transient Objects (ePESSTO+; \citealt{Smartt2015}) as a SLSN-I (see Sect.~\ref{sec:class}) on November 9, 2020 \citep{Ihanec2020}. An image of the field showing the host galaxy from the Legacy Survey \citep{Dey2018a}, as well as an image showing the SN near peak, are shown in Fig.~\ref{fig:2020zbf_image}.

We adopted a redshift of $z = 0.1947$ (see Sect.~\ref{sec:class}) and computed the distance modulus to be $39.96$~mag. In order to compute the Milky Way (MW) extinction, we used the dust extinction model of \citet{Fitzpatric1999} based on $R_{V} = 3.1$ and $E(B-V) = 0.014$~mag \citep{Schlafly2011}. As for the host galaxy extinction, we find that the host of SN\,2020zbf is a faint, blue dwarf galaxy, quite typical of SLSN-I host galaxies (Sect.~\ref{sec:host}; \citealt{Lunnan2014,Perley2016,Schulze2018}). The host galaxy analysis supports moderate host extinction [$E(B-V)_{\rm host} = 0.22^{+0.20}_{-0.22}~{\rm mag}$]. However, given that the host properties are consistent with no extinction within the uncertainties, we  
did not apply any host galaxy extinction correction to the light curves. The estimated epoch of maximum light is November 11, 2020, MJD = $59\,164.8$ (see Sect.~\ref{light_curve_properties}).

\subsection{Photometry} \label{sec:photometry}

\subsubsection{ATLAS} \label{atlas}

ATLAS is a wide-field survey consisting of four telescopes that scan the whole sky with daily cadence \citep{Tonry2018}. ATLAS observes in two wide filters, the cyan ($c$) and orange ($o$) down to a limiting magnitude of $\sim 19.7$ mag \citep{Tonry2011} and the data are processed using the pipeline described in \cite{Stalder2017}.

We retrieved forced photometry from the ATLAS forced photometry server\footnote{\url{https://fallingstar-data.com/forcedphot/}} \citep{Tonry2018,Smith2020,Shingles2021} for both $c$ and $o$ filters. We computed the weighted average of the fluxes of the observations on nightly cadence. We performed a quality cut of 3$\sigma$ in the resulting flux of each night for each filter and converted them to the AB magnitude system. The resulting data span from $-22$ to $+68$ rest-frame days post maximum light and the observed photometry is listed in Table~\ref{tab:phot_data}.

\subsubsection{Las Cumbres Observatory} \label{lco}

SN\,2020zbf was monitored by ePESSTO+ between November 2020 and December 2021 using the Las Cumbres Observatory (LCO) in the $BgVriz$ filters. The data were collected using the 1-m telescopes on South African Astronomical Observatory, Cerro Tololo Inter-american Observatory and Siding Spring Observatory. Reference images to perform image subtraction were taken
in September 2022.

We performed photometry using the AUTOmated Photometry of Transients (AUTOPhoT\footnote{\url{https://github.com/Astro-Sean/autophot}}) pipeline developed by \cite{Brennan2022}. AUTOPhoT removes host galaxy contamination through image subtraction using the HOTPANTS \citep{Becker2015} software.
The instrumental magnitude of the SN is measured through point-spread function fitting and the zero point in each image is calibrated with stars from the Legacy Survey \citep{Dey2018a} and SkyMapper Southern \citep{Onken2019} catalogs.
The LCO light curve covers the range from 0 to 180 rest-frame days post maximum light. For nights with multiple exposures, we computed the weighted average. We do not discuss the $z$-band photometry because of the poor quality of these images. The final photometry is listed in Table~\ref{tab:phot_data}.

\subsubsection{Neil Gehrels \textit{Swift} Observatory} \label{uvot}

SN\,2020zbf was observed with the UV/Optical Telescope (UVOT; \citealt{Roming2005}) on the \textit{ Neil Gehrels Swift} Observatory \citep{Gehrels2004} in all six filters, ranging from ultraviolet (UV) to visible wavelengths. The UVOT photometry is retrieved from the NASA \textit{Swift} Data Archive\footnote{\url{https://heasarc.gsfc.nasa.gov/cgi-bin/W3Browse/swift.pl}} and processed using UVOT data analysis software HEASoft version 6.19\footnote{\url{https://heasarc.gsfc.nasa.gov/}}. 
The reduction of the images is achieved by extracting the source counts from the images within a radius of 3 arcseconds and the background was estimated using a radius of 48 arcseconds. We used the \textit{Swift} tool \texttt{UVOTSOURCE} to extract the photometry using the zero points from \cite{Breeveld2011} and the calibration files from September 2020. 

The four UVOT epochs cover the range $+12$ to $+24$ rest-frame days past maximum, in all six UVOT filters. Since we have LCO $B$- and $V$-band data with better coverage, we omitted UVOT $B$- and $V$-band data from further analysis. The photometry is listed in Table~\ref{tab:phot_data}.

\subsection{Host galaxy photometry}

We retrieved the science-ready images from the DESI Legacy Imaging Surveys \citep{Dey2018a} Data Release (DR) 10 and complemented the dataset with archival $y$-band observations from the Dark Energy Survey DR 1 \citep{Abbott2018a}. The photometry was extracted with the aperture-photometry tool presented by \citet{Schulze2018}\footnote{\url{https://github.com/steveschulze/Photometry}}. Table~\ref{tab:host_phot} summarizes all measurements.

\begin{table}
\centering
\caption{Host galaxy photometry.}\label{tab:host_phot}
\begin{tabular}{cc}

\hline
\hline
Instrument/Filter & Brightness\\
                  & (mag)\\
\hline
LS/$g$        &$ 22.58  \pm 0.03$\\
LS/$r$        &$ 22.02  \pm 0.03$\\
LS/$i$        &$ 21.85  \pm 0.05$\\
LS/$z$        &$ 21.69  \pm 0.06$\\
DES/$y$       &$ 21.56  \pm 0.24$\\
\hline
\end{tabular}
\tablefoot{All magnitudes are reported in the AB system and are not corrected for MW extinction.}
\end{table}

\subsection{Spectroscopy}

We obtained six low-resolution spectra of SN\,2020zbf between November 9, 2020, and January 19, 2021, with the ESO Faint Object Spectrograph and Camera 2 \citep[EFOSC2;][]{Buzzoni1984} on the 3.58m ESO New Technology Telescope (NTT) at the La Silla Observatory in Chile under the ePESSTO+ program \citep{Smartt2015}. We complemented this dataset with one medium-resolution spectrum on November 18, 2020, with the X-shooter spectrograph \citep{Vernet2011} on the ESO Very Large Telescope (VLT) on Paranal, Chile. The spectral log is presented in Table~\ref{tab:spectra}.

The NTT spectra were reduced with the \texttt{PESSTO}\footnote{\url{https://github.com/svalenti/pessto}} pipeline. The observations were performed with grisms \#11, \#13, and \#16 using a 1\farcs0
wide slit. The integration times varied between $900$ and $5400$ s. The spectrum taken on December 8, 2020, is excluded from the analysis due to the poor signal-to-noise ratio (S/N).

The X-shooter observations were performed in nodding mode using 1\farcs0, 0\farcs9, 0\farcs9 wide slits for the UV, visible (VIS), and near-infrared (NIR) arms, respectively and were reduced using the ESO X-shooter pipeline. The procedure is the following; first the removal of cosmic-rays is done using the tool \texttt{astroscrappy}\footnote{\url{https://github.com/astropy/astroscrappy}}, based on the algorithm of \citet{vanDokkum2001a}, then the data were processed with the X-shooter pipeline v3.6.3 and the ESO workflow engine ESOReflex \citep{Goldoni2006a, Modigliani2010} and finally telluric absorption features in the VIS arm were removed with the Molecfit version 4.3.1 \citep{Smette2015a, Kausch2015a}. The wavelength calibration of all spectra was adjusted to account for barycentric motion. The spectra of the individual arms were stitched together by averaging the overlap regions.

\begin{figure*}[h]
   \centering
   \includegraphics[width=0.9\textwidth]{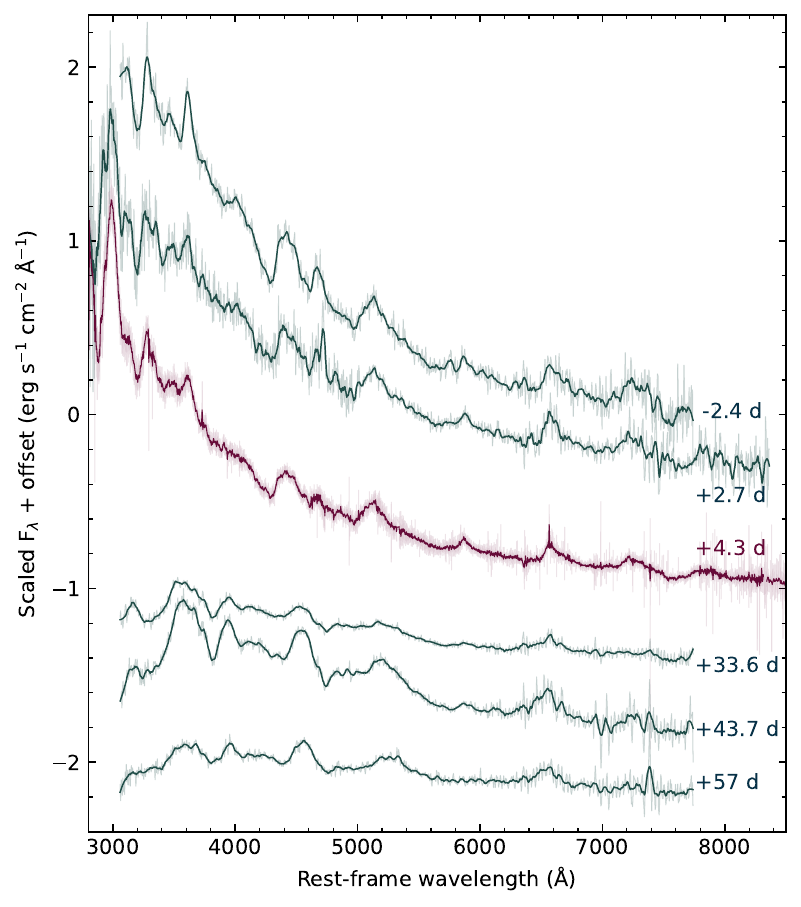}
   \caption{Spectral sequence of SN\,2020zbf from $-2.4$ to $+57$ rest-frame days. We highlight the X-shooter spectrum in purple. An offset in flux was applied for illustration purposes. The spectra are corrected for the MW extinction and are smoothed using a Savitzky-Golay filter. The original data are presented in lighter colors. See Sect.~\ref{lineID} for details on line identification.}
              \label{Spec_evolution}%
    \end{figure*}

Each spectrum was flux calibrated against standard stars. The spectral evolution from $-2.4$ to $+57$ rest-frame days past maximum brightness 
are depicted in Fig.~\ref{Spec_evolution}. All the spectra are uploaded on the WISeREP\footnote{\url{https://www.wiserep.org}} archive \citep{Yaron2012}.

\section{Light curve analysis} \label{sec:lc_analysis}

 We estimated the absolute magnitudes in each filter using the following expression:

\begin{equation} \label{abs_mag}
    M = m - \mu  - A_{\rm MW} - K_{\rm corr}
,\end{equation}

\begin{figure*}[ht]
   \centering
   \includegraphics[width=1\textwidth]{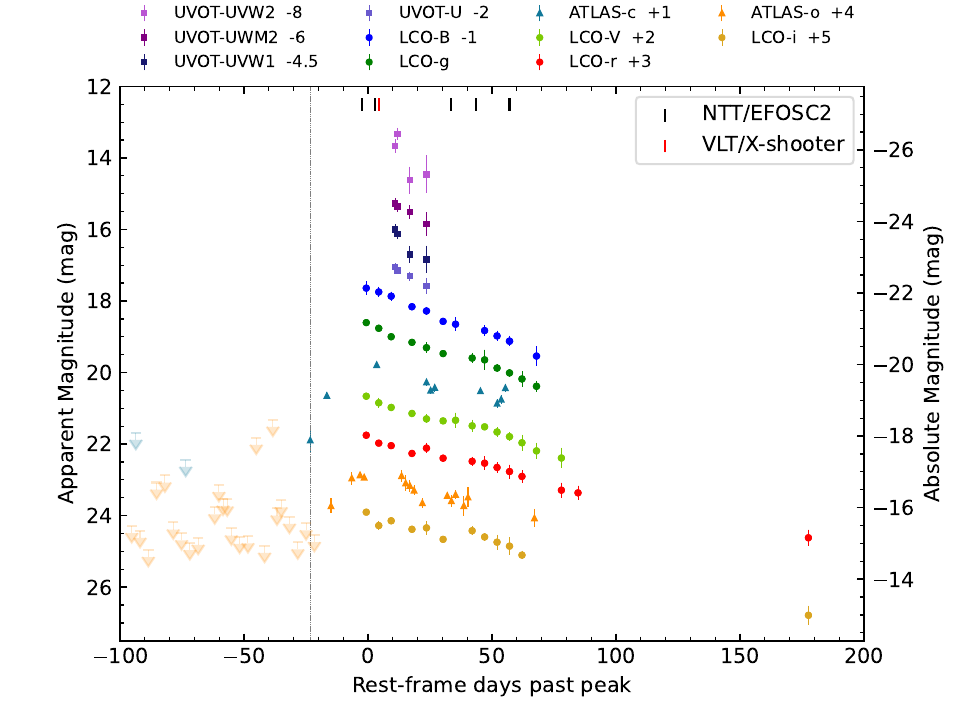}
   \caption{UV and optical light curves of SN\,2020zbf. The magnitudes are corrected for MW extinction and cosmological K-correction. Upper limits are presented as downward-pointing triangles in a lighter shade. The phase of the first detection is marked with a vertical dashed line and the epochs of the spectra are marked as thick lines at the top of the figure. The zero value on the x-axis is with respect to the g-band maximum (MJD 59164.8).}
              \label{LC}
\end{figure*}

\noindent where $m$ is the apparent magnitude, $\mu$ is the distance modulus, $A_{\rm MW}$ is the extinction caused by the MW and the last term is associated with the K-correction. The K-correction relates the photometric bandpasses in the rest frame and observer frame. It can be separated into two terms; the first term corrects for the redshift and the second term also for the shape of the spectrum \citep{Hogg2002}. In this case, we considered only the first term, $-2.5\log(1+z)$, which is a good approximation for the total K-correction as shown in \citet{Chen2023a}. We estimate $K_{\rm corr}= -0.19$~mag for all bands and epochs. The multiband light curve in apparent and absolute magnitude systems are shown in Fig.~\ref{LC}.

\subsection{Time of first light}

The rising part of the light curve was only observed with ATLAS, since LCO follow-up was triggered only after the SN was classified near peak light. Figure~\ref{LC} shows the most recent upper limits in the ATLAS $c$ and $o$ filters before the first detections (from forced photometry) at MJD~$59137.5$ and MJD~$59147.3$, respectively. Initially, we fit both the $c$ and $o$ filters separately to calculate the time of first light. However, we find that the estimated best-fit time of first light in the $o$ band is later than the first $c$-band detection. This can be understood from Fig.~\ref{LC}, since the last non-detection in the $o$ band (20.54~mag) is also after the first $c$-band detection. We therefore used the bluer $c$ band for the calculation of first light, despite the $o$ band being better sampled. The contemporaneous detection in the $c$ band and the non-detection in the $o$ band sets a limit on the color at this time to $c - o < 0.3~{\rm mag}$.

Following \cite{Miller2020}, we fit a Heaviside step function multiplied by a power law to simultaneously fit the pre-explosion baseline and the rising light curve in the ATLAS $c$ filter. Using the \texttt{PYTHON} module \texttt{emcee} \citep{Foreman2013} the power-law index $\alpha_{c}$ is estimated to be $0.68^{+0.15}_{-0.14}$ and the time of first light to be MJD $59135.4^{+1.3}_{-2.1}$. We note that these error bars only account for the statistical errors in the fit and not for any systematic errors associated with the method chosen. The approach in \cite{Miller2020} is based on the modeling of a different type of SN and the uncertainty in the explosion date in SN\,2020zbf may be larger due to qualitative differences in the rise of SLSNe-I \citep{Nicholl2016b}.

\subsection{Peak magnitude, timescales, and color evolution}  \label{light_curve_properties}
\begin{figure}[h]
   \centering
   \includegraphics[width=0.5\textwidth]{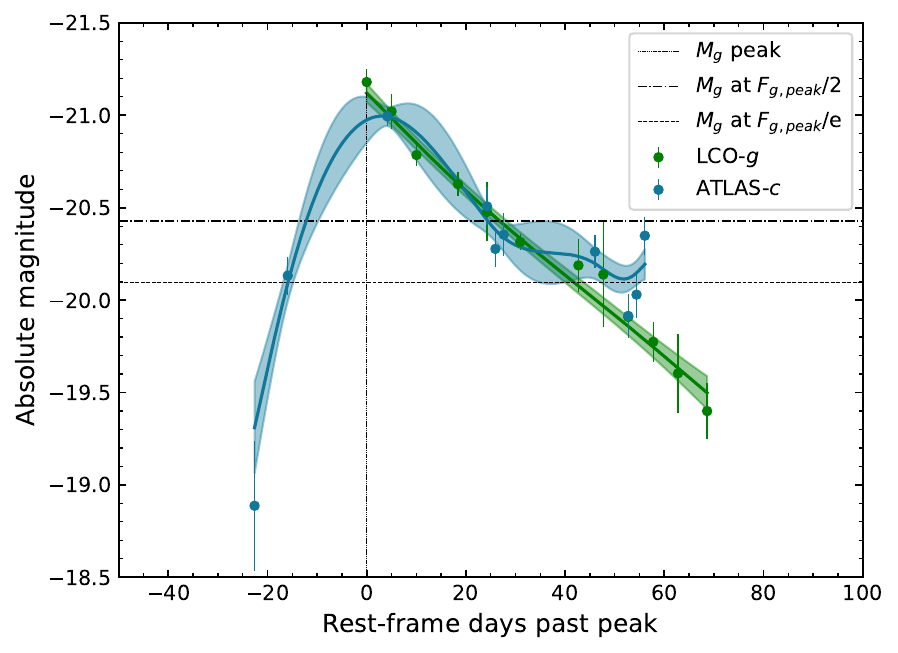}
   \caption{Gaussian process interpolation of ATLAS $c$- and LCO $g$-band light curves. The peak magnitude in the $g$ band, M$_{g}$, is indicated with the vertical line and the various rise and decline times with the horizontal lines.}
              \label{gp}%
    \end{figure}

To estimate the epoch of the maximum light as well as various light-curve timescales, we interpolated the $c$- and the $g$-band light curves. We applied the method from \cite{Angus2019} for the light-curve interpolation and fit a Gaussian process (GP) regression. To do this, we utilized the \texttt{PYTHON} package \texttt{GEORGE} \citep{Ambikasaran2015} with a Matern 3/2 kernel.

The $c$- and $g$-band photometric data with the resulting interpolations are shown in Fig.~\ref{gp}. The $g$-band light curve is already declining by the first observation, and we took the first data point as a lower limit on the $g$-band peak absolute magnitude: $M_{\rm g}$ is $-21.18 \pm 0.07$~mag, observed at MJD $59164.8$. This in turn gives an upper limit in the rise time of $ \lesssim 26.4$ rest-frame days, including the 2.1 days statistical error on the estimated time of first light. 

The rise and decline timescales of the light curve described in \cite{Chen2023a} are determined using the $c$-band interpolated light curve and the maximum $M_{\rm g}$. The rest-frame rise time from the half maximum flux ($F_{\rm g,peak}$/2) is $12.2_{-2.4}^{+1.2}$ days and from 1/$e$ maximum flux ($F_{\rm g,peak}$/$e$) is $15.9_{-1.1}^{+1}$ days. We estimated the rest-frame decline times using the $g$-band interpolated light curve. The decline time to the half maximum flux is $26.86_{-1.44}^{+1.44}$ days and to the 1/$e$ maximum flux is $41.9_{-1.9}^{+2}$ days. These are shown in gray lines in Fig.~\ref{gp}.

\begin{figure}[h]
   \centering
  \includegraphics[width=0.5\textwidth]{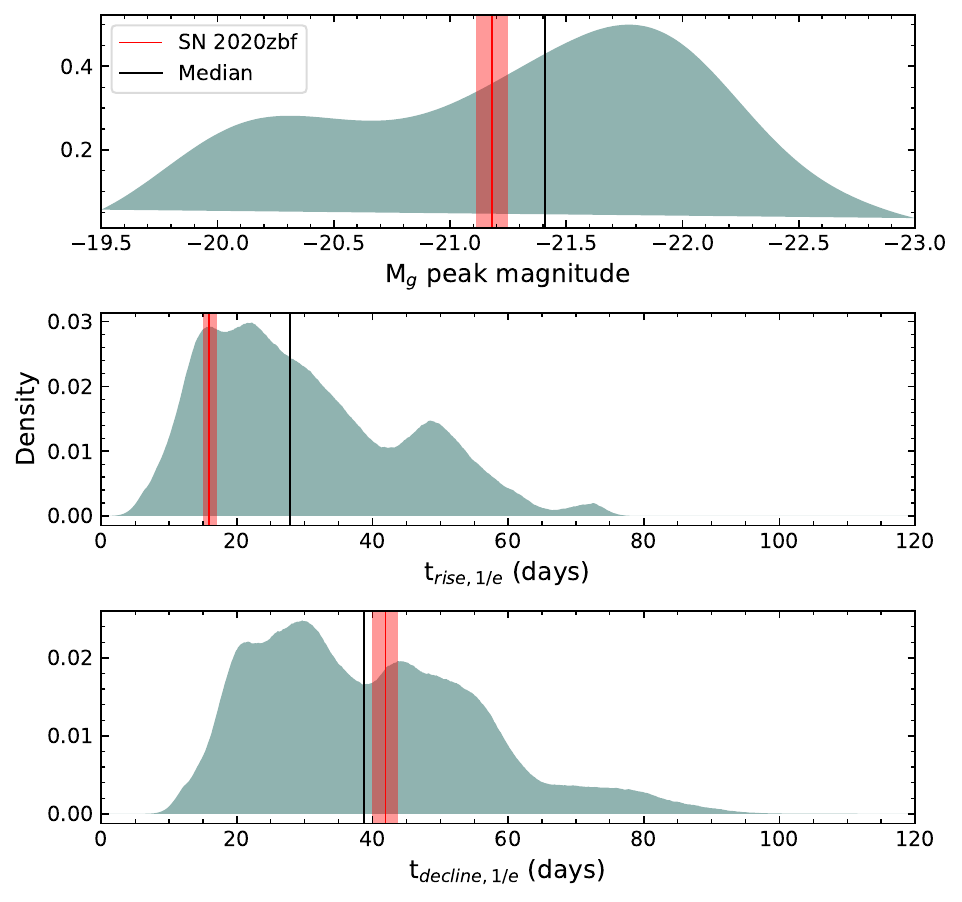}
   \caption{Comparison of the photometric properties of SN\,2020zbf with the ZTF SLSN-I sample \citep{Chen2023a}. Top: KDE distribution of the M$_{g}$ peak magnitudes for 78 ZTF SLSNe-I. Middle: KDE plot of the $e$-folding rise time for 69 ZTF SLSNe-I. Bottom: $e$-folding decline time distribution for 54 ZTF SLSNe-I. The vertical red line along with the errors (shaded red regions) illustrate the position of SN\,2020zbf and the black vertical lines the median values.}
      \label{phot_comp}%
   \end{figure}

In Fig.~\ref{phot_comp}, we put the light curve properties of SN\,2020zbf in the context of the homogeneous Zwicky Transient Facility (ZTF; \citealt{Bellm2019}) SLSN-I sample from \cite{Chen2023a}. This paper studied the UV and optical photometric properties of 78 H-poor SLSNe-I. In the three different panels, we show the kernel density estimates (KDEs) of the ZTF sample, which are an outcome of a Monte Carlo simulation accounting for the asymmetric errors, and indicate by the red vertical lines the measurements for SN\,2020zbf. The peak absolute magnitude is fairly average, being slightly fainter than the median value of the SLSNe-I. In contrast, the rise time of SN\,2020zbf is among the fastest seen for SLSNe-I, whereas the decline is again rather average.

\begin{figure}[h]
   \centering
  \includegraphics[width=0.5\textwidth]{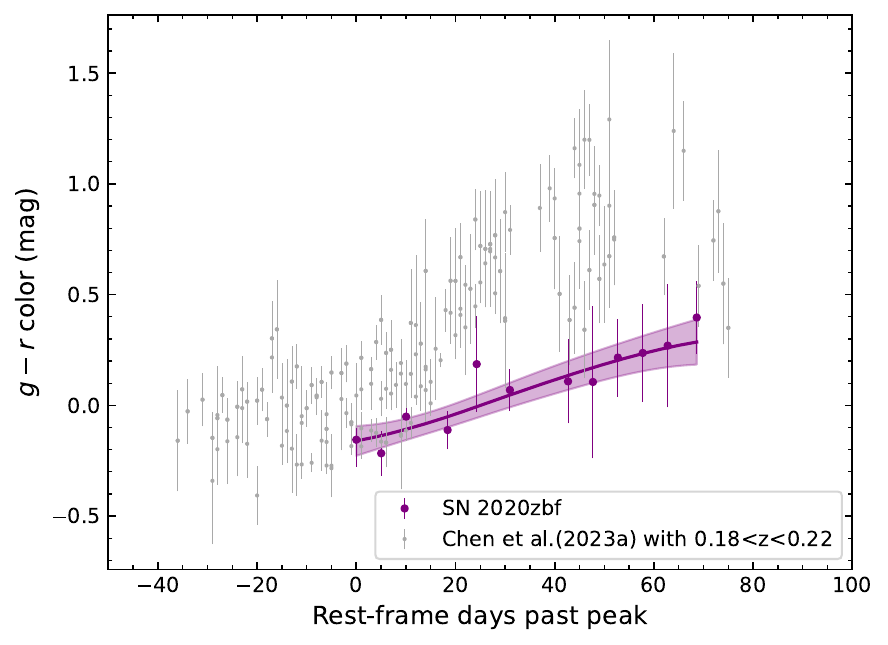}
   \caption{ $g-r$ color evolution of SN\,2020zbf along with the GP interpolation in light purple. For comparison, the ZTF $g-r$ color evolution of the \cite{Chen2023a} SLSN sample with $ 0.18$ $<$ z $<$ $0.22$ is presented in gray.}
      \label{color_evolution}%
   \end{figure}

To construct the $g-r$ color evolution of SN\,2020zbf, we used the $g$- and $r$-band interpolated light curves and plot the results in Fig.~\ref{color_evolution}. For comparison, we present the reddening corrected $g-r$ colors of the SLSNe-I from the ZTF sample of \cite{Chen2023a} with redshifts within $\pm 0.02$ of SN\,2020zbf's redshift (in order to facilitate comparison at similar effective wavelengths). Although the $g-r$ color evolution of SN\,2020zbf follows the general trend of the ZTF sample, by getting redder over time, it evolves more slowly than other SLSNe-I showing a consistently bluer color.

\subsection{Photospheric temperature and radius} \label{BB}

\begin{figure}[h]
   \centering
   \includegraphics[width=0.5\textwidth]{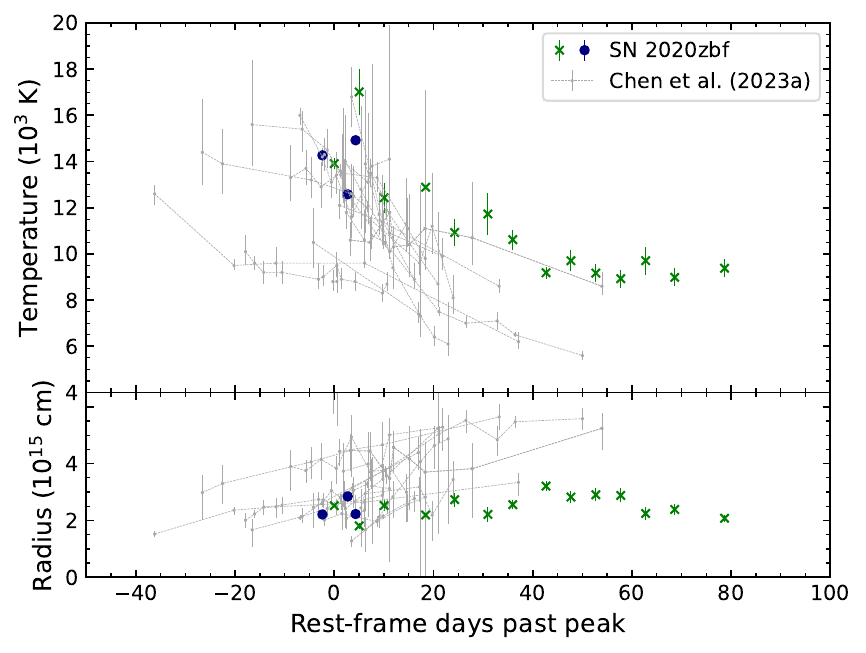}
   \caption{Blackbody temperatures and radii of SN\,2020zbf. Top: Temperature evolution of SN\,2020zbf derived from the blackbody fits of the photometric (green) and the spectroscopic (blue) data. The gray background points present the temperature evolution of the ZTF sample \citep{Chen2023a}. Bottom: Blackbody radius evolution of SN\,2020zbf using photometry (green) and the early spectra (blue).}
              \label{temperature}%
    \end{figure}

We interpolated all the UVOT and LCO light curves using the GP method described in Sect.~\ref{light_curve_properties} and extracted the magnitudes using the $V$-band epochs as reference since it has the most observed epochs. We excluded the ATLAS filters because they are significantly broader than the LCO filters. We constructed the spectral energy distributions (SEDs) by calculating the spectral luminosities $L_{\rm \lambda}$ for each band, at each of the 14 past-peak epochs, and fit a blackbody utilizing the \texttt{scipy.optimize.curvefit}\footnote{\url{https://docs.scipy.org/doc/scipy/reference/generated/scipy.optimize.curve_fit.html}} module. Due to line blanketing \citep{Yan2017b}, we excluded the UVOT data from the blackbody fits. The resulting photospheric $BgVri$ temperature and radius evolution are plotted in Fig.~\ref{temperature}.

To check whether there is consistency with the spectral measurements, we estimated the temperature and the radius by fitting a blackbody to the spectra taken in the early photospheric phase. We first absolute-calibrated the spectra with the 
photometric data before template subtraction and then corrected them for the MW extinction. The results are shown in Fig.~\ref{temperature} along with the SLSNe-I from the \cite{Chen2023a} sample in light gray. We chose to compare with the ZTF SLSNe-I, which are characterized as normal events by \cite{Chen2023a}, excluding the objects with fewer than two epochs as well as three objects marked in \cite{Chen2023a} as extraordinary events. Overall, the temperature evolution of SN\,2020zbf is comparable to those of the ZTF sample but the temperatures are higher than for the bulk of the population, which is in agreement with the color evolution in Fig.~\ref{color_evolution}.
The radius of the photosphere shows a rise trend up to $\sim 45$ days, which is consistent with what is seen in the ZTF sample \citep{Chen2023a}. After 45 days, the photospheric radius seems to decline.

\subsection{Bolometric light curve} \label{bolometric_properties}

\begin{figure}[h]
   \centering
   \includegraphics[width=0.5\textwidth]{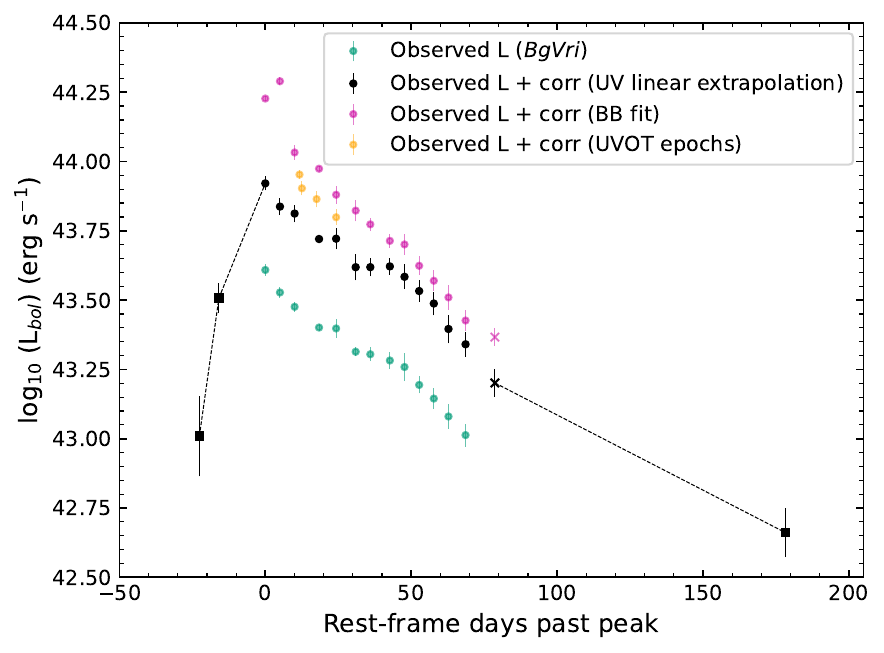}

   \caption{Bolometric light curve of SN\,2020zbf with different corrections applied. The circles correspond to the derived luminosities using all LCO filters, the crosses
   are the $Vri$ bands and the square symbols illustrate the bolometric luminosity assuming the same bolometric correction as the epochs with multiband data. The green light curve corresponds to the integrated observed $BgVri$ flux and constitutes a lower limit on the total bolometric luminosity. The black symbols include a correction to the NIR and the UV using the UV linear extrapolation method \citep{Lyman2014}, whereas the purple light curve considers a UV correction by integrating the blackbody fit. The orange data points show the observed bolometric luminosity including the NIR correction if we consider also the UVOT data. The errors represent statistical errors. For our analysis we used the black curve, and the dashed lines connect the correction data points for illustration purposes.}
              \label{BC}%
    \end{figure}

To construct a bolometric light curve, we started by integrating the observed SED ($BgVri$) at the epochs we have LCO data. The result is shown as green points in Fig.~\ref{BC}, and constitutes a strict lower limit on the bolometric luminosity from the observed flux over the optical bands only. For a better estimate on the total bolometric luminosity, we considered different methods to account for the missing flux. For the NIR correction, we fit a blackbody to the LCO data as before and integrated the blackbody tail up to 24\,400~\AA, beyond which the contribution to the bolometric light curve is negligible in the photospheric phase \citep[$\sim$ 1\%;][]{Ergon2013}.

For the UV correction, which constitutes a significant fraction of the bolometric flux at early times when the temperature is high, we used two different methods for comparison. First, we considered a blackbody method, integrating the same blackbody fit from 0~\AA\ to the $B$ band, and adding up the UV, observed optical, and NIR flux. This is shown as the purple curve in Fig.~\ref{BC}, and can be considered an upper limit on the bolometric luminosity, since it does not take into account UV line blanketing. To better capture this effect, we finally linearly extrapolated the SED from the $B$ band to 2000~\AA\ (where $L_{\lambda}$ is assumed to be zero), following \citet{Lyman2014}; the pseudo-bolometric light curve using this UV correction is shown as black points in Fig.~\ref{BC}. Near peak, the linear extrapolation method adds $\sim$ 0.31~dex to the observed flux while the blackbody method adds $\sim$ 0.62~dex. In later epochs, when the UV emission is small, the two luminosities have consistent values.

We tested the validity of our UV corrections against the four epochs for which we have UVOT data. We integrated the observed SED using the $BgVri$ and UVOT filters and added the NIR flux we calculated above. The four-epoch bolometric light curve is shown in orange in Fig.~\ref{BC}. The data points are placed between the UV linear extrapolation method \citep{Lyman2014} and the blackbody fit corrected light curves, indicating that the flux is overestimated by $\sim$ 0.1~dex when integrating the full blackbody in the UV and underestimated by $\sim$ 0.1~dex when using the UV linear correction. Since we know that SLSNe-I do have significant UV absorption \citep{Yan2017b}, we used the linear correction for the rest of the analysis but note that this also somewhat underestimates the total flux. The peak bolometric luminosity  $L_{\rm bol}^{\rm peak}$  is estimated to be $\gtrsim$ 8.3~$\times$~10$^{43}$~erg~s$^{-1}$.

At our earliest and latest epochs, we did not have the SED information to perform the same analysis as above. In order to include these points in the bolometric light curve (for purposes of integrating the total energy), we instead assumed a constant bolometric correction. For the data on the rise, we used the $c$ band and used the same ratio of the $c$-band flux to total flux that we measured at the first epoch with multiband data. Since we expect the temperature to be higher at these earlier epochs, this approximation will be a lower limit. Similarly, our final epoch only has $r$ and $i$ measurements, and we scaled the bolometric luminosity to the latest multiband measurement. These points are shown as square symbols in Fig.~\ref{BC}. Integrating the pseudo-bolometric light curve, including these early and late measurements, we find the total radiated energy of SN\,2020zbf to be $E_{\rm rad} \gtrsim 4.68 \pm 0.14 \times 10^{50}~{\rm erg}$, consistent with what we would expect for SLSNe-I \citep{Quimby2011}.

\section{Spectral analysis} \label{sec:spectra}

\subsection{Host galaxy redshift and spectral classification} \label{sec:class}

\begin{figure}[h]
   \centering
   \includegraphics[width=0.5\textwidth]{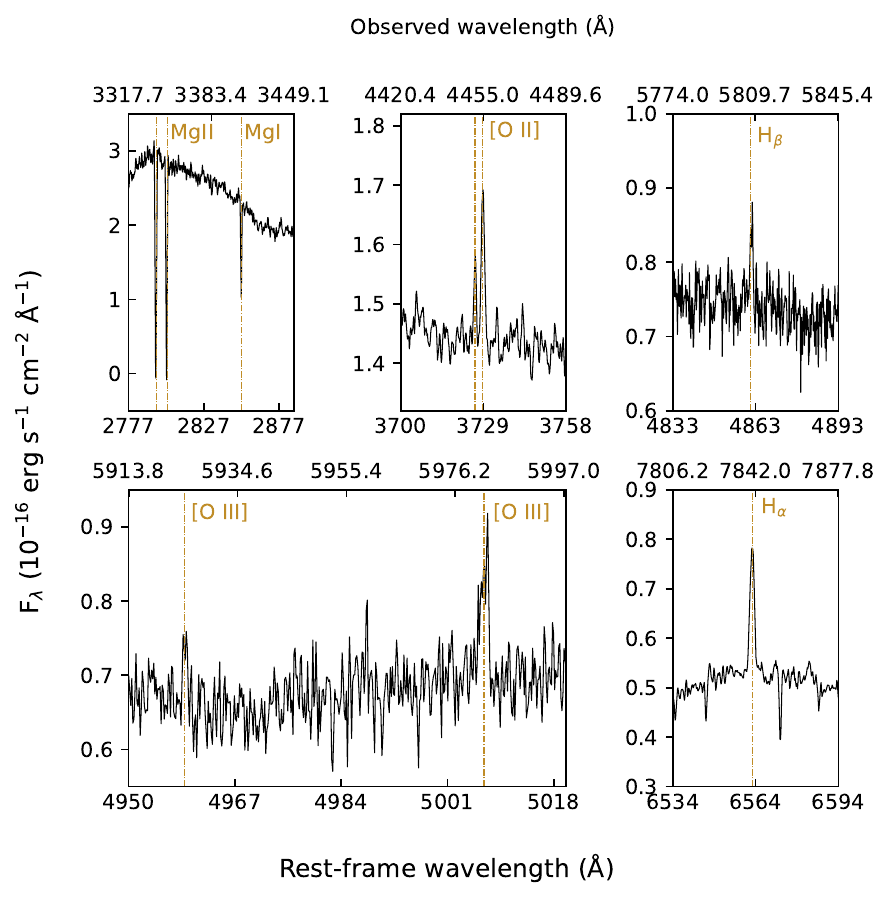}
   \caption{Host galaxy absorption and emission lines in the X-shooter spectrum of SN\,2020zbf $+4.3$ days after peak. The absorption and emission lines give a consistent host redshift of $z=0.1947$.}        
   \label{fig:hostlines}
    \end{figure}

The first spectrum of SN\,2020zbf was obtained on November 9, 2020, and and was classified by \cite{Ihanec2020} as a SLSN-Ic at $z = 0.35$ using the cross matching tool SNID \citep{Blondin2007}; it was found to be a good match to the SLSN-I PTF12dam \citep{Nicholl2013}. Based on this classification we triggered our X-shooter program; the higher-quality spectrum taken on November 18, 2020, reveals narrow emission and absorption lines consistent with a galaxy at a redshift of $z = 0.195$ \citep{Lunnan2021}. 
We reexamined the spectrum and identified emission lines from the interstellar medium and \ion{H}{ii} regions in the host galaxy at a common redshift of $z = 0.1947 \pm 0.0001$. Figure~\ref{fig:hostlines} shows the host lines that were used for the redshift determination: the galaxy's narrow \ion{Mg}{II} doublet $\lambda\large2796, 2803$, the \ion{Mg}{I} $\lambda2852$, the [\ion{O}{II}] doublet $\lambda\lambda3727, 3729$, the narrow H${\alpha}$ $\lambda6563$, the  H$\beta$ $\lambda 4861,$ and the forbidden [\ion{O}{III}] doublet $\lambda\lambda4959,5007$. The initial redshift misclassification highlights both the peculiarity of SN\,2020zbf in comparison with typical SLSNe-I, as well as the challenge of classifying unusual events based on lower-quality data.

Having established a robust redshift, we tentatively identified the two absorption features around 4500~\AA\ as \ion{O}{II}, supporting a SLSN-I spectroscopic classification \citep[e.g.,][]{Quimby2018}, although the velocities would be quite low (see Sect.~\ref{lineID}). Even without the \ion{O}{II} line identification, we argue that SN\,2020zbf is best classified as a SLSN-Ic. The absolute magnitude of SN\,2020zbf in the $g$ band is $-21.18$~mag (see Sect.~\ref{light_curve_properties}), which is well within the SLSN brightness regime \citep{GalYam2012,DeCia2018,Inserra2018,Angus2019,Chen2023a}. There is no evidence of H lines in the SN spectrum (the feature at 6578~\AA\ is more likely to be \ion{C}{II}; see Sect.~\ref{lineID}) and also no obvious He lines. Finally, as the ejecta cool, the spectrum evolves to look like a typical SN-Ic.

\subsection{Line identification}  \label{lineID}
\begin{figure*}[h]
   \centering
   \includegraphics[width=1\textwidth]{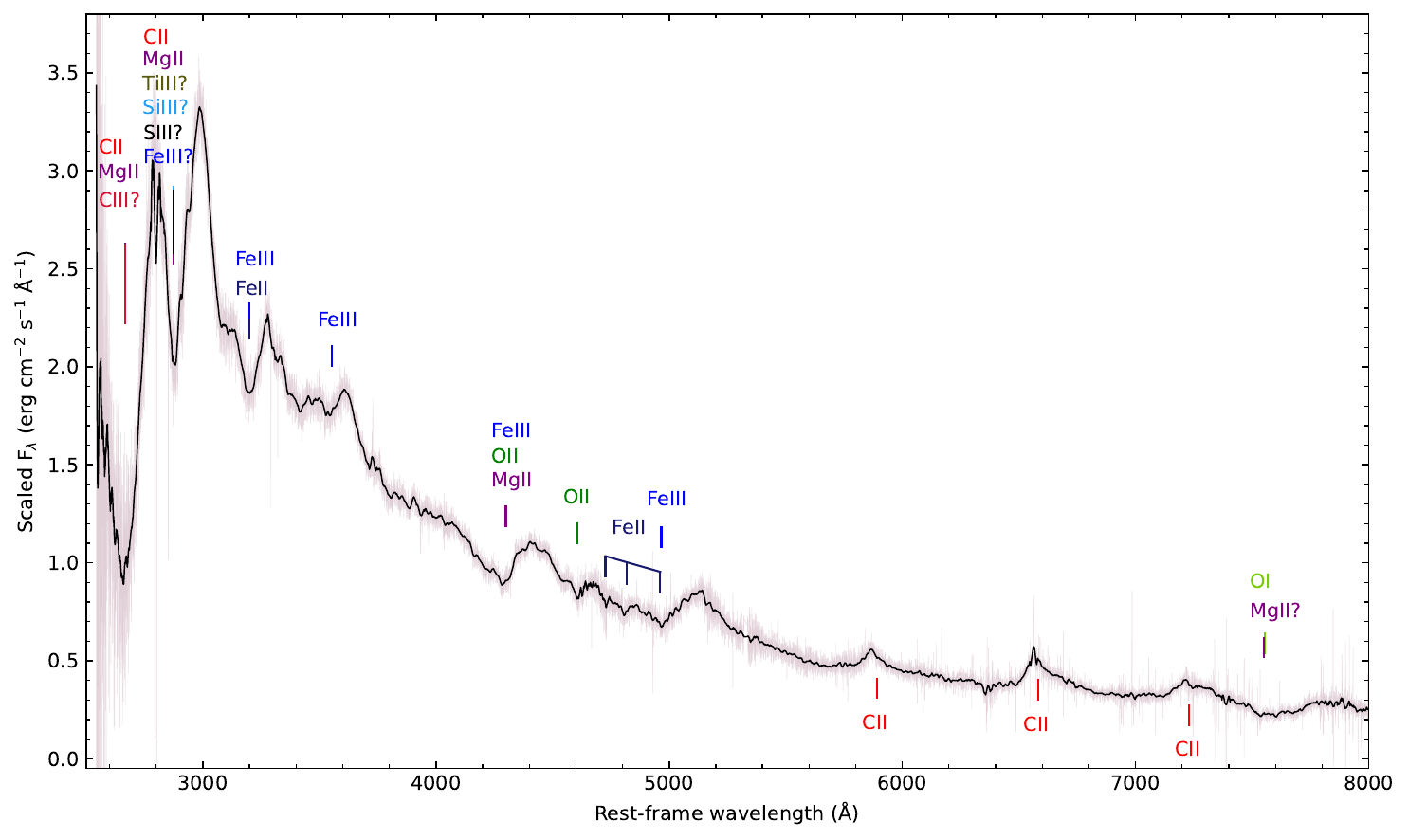}
   \caption{X-shooter spectrum of SN\,2020zbf at $+4.3$ days after maximum light. The spectrum is corrected for MW extinction and is smoothed using a Savitzky-Golay filter. The original spectrum is shown in lighter colors. The most conspicuous features are labeled. Uncertain line identifications are denoted with question marks.
 The ions beneath the spectrum are shown at $v$ = 0, whilst those above have been shifted to match the absorption component. }
              \label{xshooter}%
    \end{figure*}

\begin{figure}[h]
   \centering
   \includegraphics[width=0.5\textwidth]{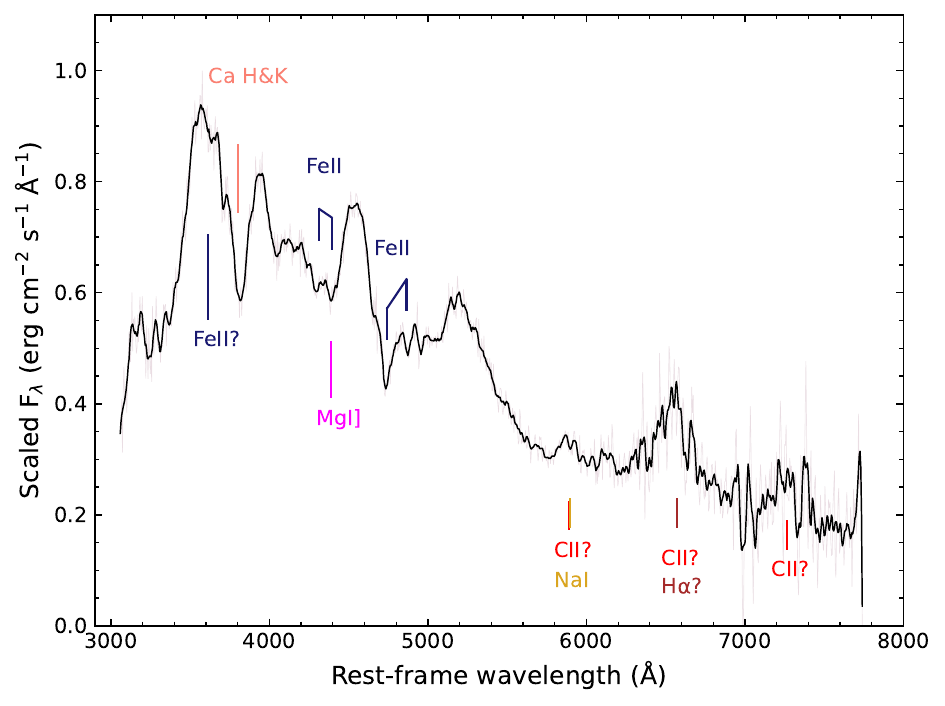}
   \caption{NTT-EFOSC2 spectrum of SN\,2020zbf at $+43.7$ days after maximum light. The spectrum is corrected for MW extinction and is smoothed using a Savitzky-Golay filter. The original spectrum is shown in lighter colors. The colors depict the various elements and the question marks indicate that their presence in the spectrum is unclear. The ions beneath the spectrum are shown at the rest-frame wavelength, whilst those above have been shifted to match the absorption component.}
              \label{ntt}%
\end{figure}

Line identification of SN\,2020zbf was performed using the medium-resolution X-shooter spectrum at $+4.3$ days and the low-resolution NTT-EFOSC2 spectrum at $+43.7$ days after maximum brightness; these spectra have the highest S/N. The identification was done by comparing with other SLSNe from the literature \citep{Inserra2013,Anderson2018,Quimby2018,GalYam2019a,Pursiainen2022,Tinyanont2023}, with the predictions of spectral models \citep{Dessart2012,Mazzali2016,Dessart2019}, and finally by searching the National Institute of Standards and Technology (NIST; \citealt{Kramida2022}) atomic spectra database for lines above a certain strength, similar to what was done in \citet{GalYam2019a}. Figures~\ref{xshooter} and~\ref{ntt} show the $+4.3$ and $+43.7$ days spectra, respectively, along with the most conspicuous features blueshifted by 9000 -- 12\,000~km~s$^{-1}$ (see Sect.~\ref{sec:velocity}). The \ion{C}{II} lines are shown at zero rest-frame velocity.

Early spectra of SLSNe-I are characterized mostly by the presence of strong \ion{O}{II} lines between 3500 and 5000~\AA\ \citep{Quimby2011,Mazzali2016,Quimby2018}. In the case of SN\,2020zbf, only the \ion{O}{II} $\lambda 4358$ and $\lambda 4651$  W-shape is visible with absorption troughs consistent with a velocity of 4000~km~s$^{-1}$ (see Sect.~\ref{sec:velocity}), which is relatively low for SLSNe-I (median value of 9700~km~s$^{-1}$; \citealt{Chen2023a}). The absorption at 4300~\AA\ seems stronger and broader compared to the one at 4600~\AA, indicating the presence of other ions at these wavelengths \citep{Nicholl2016a}. In particular, we associate the feature at 4300~\AA\ with a blend of \ion{O}{II} $\lambda 4358$, \ion{Fe}{III} $\lambda 4432$ and \ion{Mg}{II} $\lambda 4481$. Redward of the \ion{O}{II} $\lambda 4651$, the \ion{Fe}{II} triplet $ \lambda\lambda 4923,5018,5169$ is present, but the \ion{Fe}{II} $\lambda5169$ is likely mixed with \ion{Fe}{III} $\lambda5129$ \citep{Liu2017}.

Blueward of 3000~\AA, the UV part of the spectrum is very blended and it is hard to identify individual lines. There are two strong absorption components at 2670~\AA\ and 2880~\AA. The first trough at 2670~\AA\ is associated with \ion{Mg}{II}, \ion{C}{II,} and \ion{C}{III,} which is consistent with what is seen in other SLSNe-I \citep{Quimby2011,Lunnan2013,Howell2013,Vreeswijk2014,Yan2017b,Yan2018,Smith2018} and in spectral models \citep{Dessart2012,Mazzali2016,Dessart2019}. The second absorption at 2880~\AA\ has been observed in a number of SLSNe (iPTF\,13ajg; \cite{Vreeswijk2014}, PTF09atu and PTF12dam; \citealt{Quimby2018}) but not as strong as in SN\,2020zbf and has never been conclusively identified. According to \cite{Mazzali2016} and \cite{Quimby2018} these features are mostly attributed to \ion{Ti}{III}. \cite{Dessart2012} also suggested \ion{Fe}{III}, \ion{Si}{III,} and \ion{S}{III}. Searching the NIST library, we find that other possible contributions in this region could be \ion{C}{II} and \ion{Mg}{II}.

Between 3000 and 3600~\AA, the absorption at 3200~\AA\ could be attributed to \ion{Fe}{II} $\lambda3325$ and \ion{Fe}{III} $\lambda3305$. The feature at 3550~\AA\ is associated with \ion{Fe}{III} $\lambda3691$, but we were unable to identify the ions that may contribute to the feature at 3410~\AA. We also see an absorption feature at 7550~\AA\ that could be formed by the \ion{O}{I} triplet $\lambda\lambda7772,7774,7775$ and a small contribution of \ion{Mg}{II} $\lambda 8234$.

Figure~\ref{ntt} shows the spectrum at $+43.7$ days. The spectrum at this stage resembles that of a normal SN-Ic at maximum light  \citep{Pastorello2010,Quimby2011}. The \ion{O}{II} and \ion{Fe}{III} lines, which dominated the early spectra, have disappeared and elements from further in are revealed as the ejecta cool down. We see \ion{Ca}{II} $\lambda\lambda 3966,3934$, \ion{Mg}{I}] $\lambda 4571$ and strong \ion{Fe}{II} lines between 4000 and 5200~\AA\ blueshifted by 9000 km s$^{-1}$ (see Sect.~\ref{sec:velocity}) to match the absorption component. We cannot distinguish the \ion{O}{I} triplet $\lambda\lambda 7772,7774,7775$ due to the low S/N. At 3600~\AA, there is a broad emission component that has been observed in the SLSNe-Ic LSQ12dlf \citep{Nicholl2014,Nicholl2016b}, SN\,2007bi \citep{Young2010,Nicholl2013}, and SN\,2017egm \citep{Bose2018}. In \cite{Nicholl2014} and \cite{Bose2018} the result from spectrum synthesis code for LSQ12dlf and SN\,2017egm, respectively, shows that this feature is associated with \ion{Fe}{II} while in SN\,2007bi this line is not identified.

\subsection{\ion{C}{II} lines} \label{sec:cii}

In Fig.~\ref{xshooter}, the most prominent features in the red part of the optical spectrum are likely attributed to \ion{C}{II} $\lambda 5890$, $ \lambda 6580,$ and $\lambda  7234$. These lines have been predicted by various SLSN-I models \citep{Dessart2012,Mazzali2016,Dessart2019} and have been seen in several SLSNe-Ic (see Sect.~\ref{comparison}) but are typically much weaker than seen in SN\,2020zbf. If we consider that the broad emission component at 6580~\AA\ at $+4.3$ days spectrum is associated with H${\alpha}$ rather than \ion{C}{II}, we should be able to detect a broad H${\beta}$ emission at 4861~\AA. The region around H$\beta$ is dominated by \ion{Fe}{II} absorption, complicating the analysis, but we do not find evidence for any broad Balmer emission.

\begin{figure}[h]
   \centering
   \includegraphics[width=0.5\textwidth]{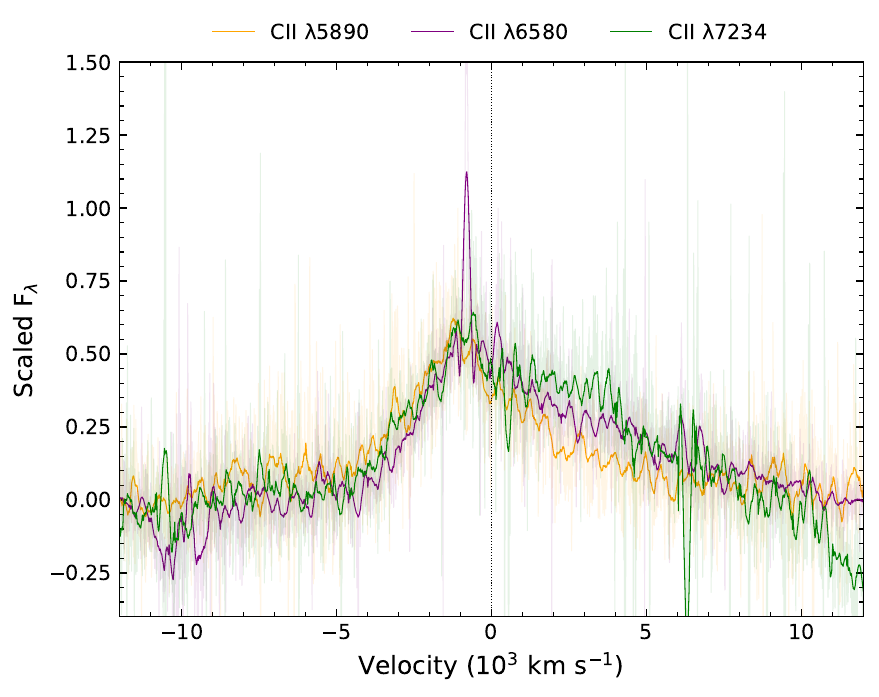}
   \caption{Three \ion{C}{II} emission lines detected in the X-shooter spectrum $+4.3$ days after peak. A local linear continuum subtraction and a normalization to the peak flux has been applied for illustration purposes. The spectra have been smoothed using a Savitzky-Golay filter while the original data are shown in a lighter color. The rest-frame wavelength of the \ion{C}{II} emission is denoted by the dashed vertical line. Note that the narrow emission component at 6580~\AA\ is the H$\alpha$ emission from the host galaxy.}
   
              \label{cii_profiles}%
\end{figure}

For comparison, the three \ion{C}{II} profiles in the X-shooter spectrum at $+4.3$ days post maximum are shown in Fig.~\ref{cii_profiles}. The emission line profiles appear to be consistent with one another
supporting the hypothesis of \ion{C}{II} at 6580~\AA\ over H$\alpha$. The \ion{C}{II} profiles show an extended tail in the red wing and their peaks are blueshifted by $\sim$ 1000~km~s$^{-1}$ relative to zero rest-frame velocity, which can be explained by multiple electron scatterings \citep{Branch2017,Jerkstrand2017}. However, the lines have a triangular shape, which could indicate a different formation process than for the rest of the emission lines in the spectrum. The \ion{C}{II} lines do not show obvious P-Cygni profiles since there is no evident absorption component.

\begin{figure}[h]
   \centering
   \includegraphics[width=0.5\textwidth]{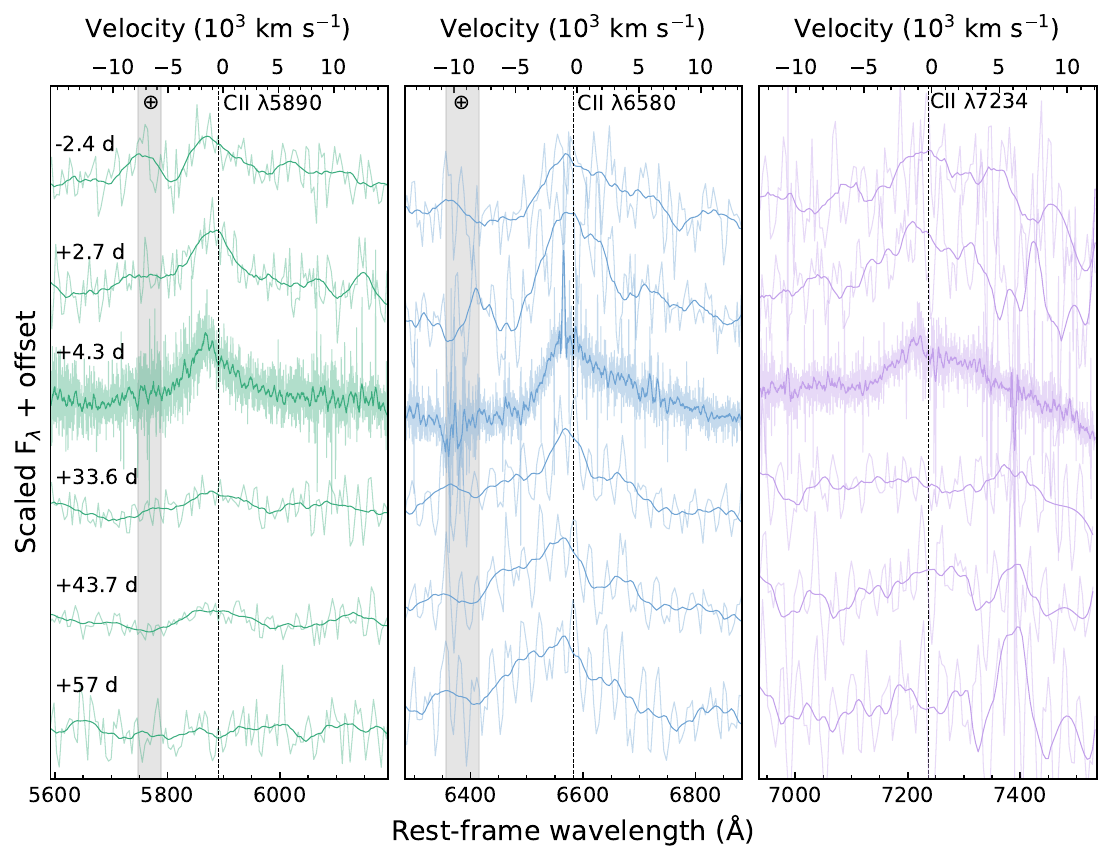}
   \caption{Evolution of the three \ion{C}{II} lines of SN\,2020zbf from $-2.4$ to $+57$ rest-frame days. The telluric absorptions are indicated with lighter gray vertical bands. The spectra have been corrected for extinction. A smoothing has been applied utilizing the Savitzky-Golay filter and the original data are shown in lighter colors. The regions corrected for telluric absorptions are indicated with lighter gray vertical lines. A local linear continuum subtraction and an offset were applied for display purposes. The rest-frame wavelengths of the \ion{C}{II} lines are illustrated with dashed vertical lines.}
    \label{cii_evolution}%
\end{figure}

Figure~\ref{cii_evolution} illustrates the evolution of the \ion{C}{II} lines. The emissions at 5890~\AA\ and 7300~\AA\ become weaker than in the early photospheric phase, whereas the component at 6580~\AA\ remains strong. The weak emission component at 5890~\AA\ can be attributed mostly to \ion{Na}{I} $\lambda\lambda5890,5896$. The most intriguing characteristic of the spectrum at $+43.7$ days is the persistent signal at around 6580 ~\AA,\ which could be attributed to the late-time appearance of H$\alpha$ (see Sect.~\ref{sec:csm_discussion}). In Fig.~\ref{cii_evolution} we note that the smoothed data at $-2.4$ days for \ion{C}{II} $\lambda 5890$ and $ \lambda 6580$ and at $+2.7$ days for \ion{C}{II} $\lambda 6580$  show an absorption component blueshifted by 5000, 8000, and 6000~km~s$^{-1}$, respectively. However, these components are not clearly seen in the original data and the velocity values are not consistent across the three \ion{C}{II} lines. In addition, any feature blueward \ion{C}{II} $\lambda 5890$ and $ \lambda 6580$ could be an artifact from the telluric removal. Thus, we cannot conclusively consider them as the absorption troughs of a P-Cygni profile. The absence of absorption troughs in \ion{C}{II} $\lambda 7234$ as well as in Fig.~\ref{cii_profiles} strengthens this argument. For the reasons stated above, we conclude that the \ion{C}{II} lines are more likely pure emission. 

\subsection{Line velocities} \label{sec:velocity}

There are two main absorption line features commonly used to derive the velocity of the photosphere from the spectra in SLSN-I. The first method, shown in \cite{Quimby2018} and \cite{GalYam2019b,GalYam2019a}, is to use the \ion{O}{II} absorption lines at 3500 -- 5000~\AA\ at early phases. As mentioned in Sect.~\ref{sec:class}, the only visible \ion{O}{II} lines in the spectra of SN\,2020zbf are the $\lambda4358$ and $\lambda4651,$ and the absorption troughs are shifted by $- 4000$~km~s$^{-1}$, which is low compared to other SLSNe-I \citep{Quimby2018,GalYam2019b}. \cite{Chen2023b} studying a sample of 77 SLSNe-I finds the median \ion{O}{II} velocity of the ZTF sample to be 9700~km~s$^{-1}$ with the distribution going down to 3000~km~s$^{-1}$. \cite{Nicholl2016b} studied a sample of 24 SLSNe-Ic and showed that the median velocity of the SLSNe-I is 10\,500 $\pm$ 3000 km s$^{-1}$. Our estimated value of 4000~km~s$^{-1}$ is lower than the median, but since it  has been observed in at least one SLSN in the \cite{Chen2023b} sample, this value is not unprecedented.

\begin{figure}[h]
   \centering
   \includegraphics[width=0.5\textwidth]{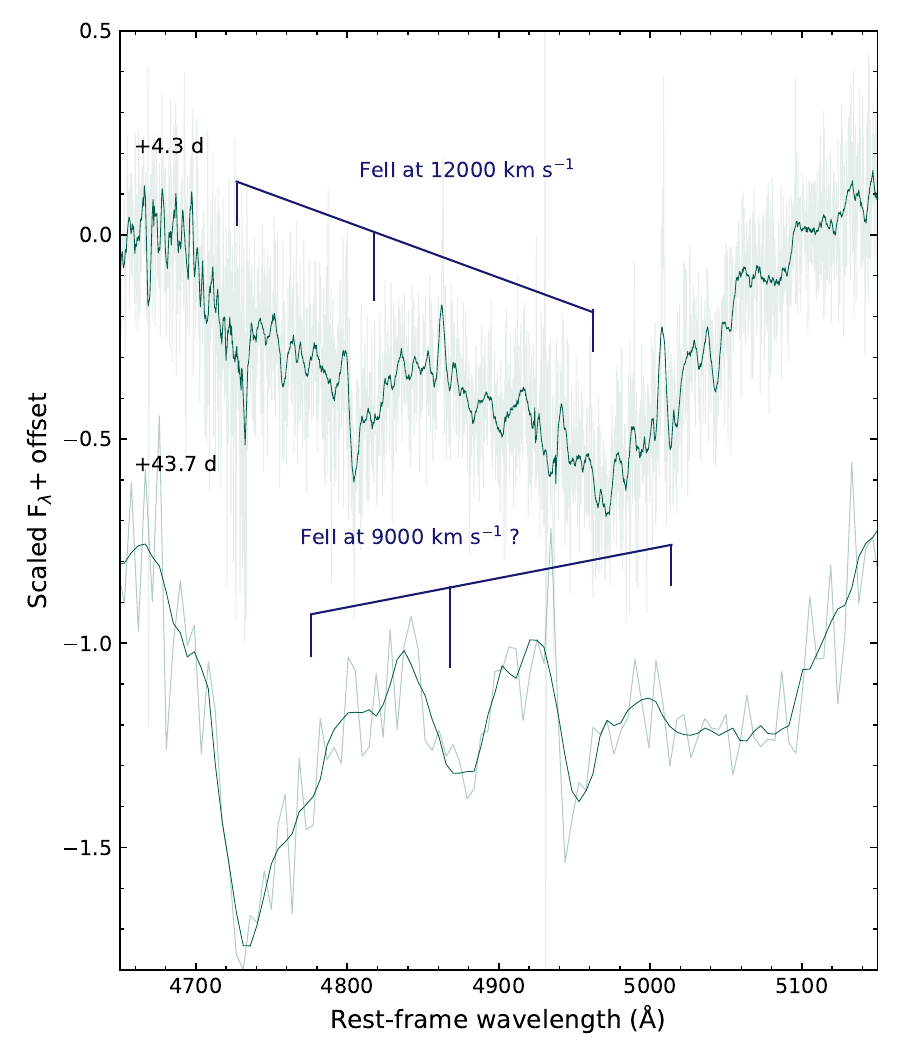}
   \caption{\ion{Fe}{II} triplet $\lambda\lambda4923,5018,5169$ region of the $+4.3$ and $+43.7$ day spectra. A linear continuum subtraction and an arbitrary offset has been applied for illustration purposes. The spectra have been smoothed using the Savitzky-Golay filter and the original data are shown in lighter colors. The absorption features that correspond to the blueshift of the \ion{Fe}{II} lines are marked in blue along with the velocity. At $+43.7$ days, the absorption feature at 4740\AA\ could imply that \ion{Fe}{II} is still at 12\,000~km~s$^{-1}$.}
              \label{feii_evolution}%
\end{figure}

The second method for measuring the photospheric velocity is to use the \ion{Fe}{II} triplet $\lambda\lambda4923,5018,5169$ as tracers \citep{Branch2002,Nicholl2016b,Modjaz2016,Liu2017}. We note that \ion{Fe}{II} $\lambda5169$ seems blended with \ion{Fe}{III} $\lambda5129$ in the hot photospheric phase and \ion{Fe}{II} $\lambda4923$ is poorly detected. Thus, we used \ion{Fe}{II} $\lambda5018$ to estimate the velocity. In Fig.~\ref{feii_evolution} a zoomed-in view of the \ion{Fe}{II} triplet region at $+4.3$ and $+43.7$ days post-maximum is shown. In the X-shooter spectrum at $+4.3$ days after peak, the marked absorption components match well with the \ion{Fe}{II} triplet shifted by $-12\,000$~km~s$^{-1}$ even though the \ion{Fe}{II} $\lambda4923$ is poorly resolved and the \ion{Fe}{II} $\lambda5169$ is blended. This velocity is consistent within the errors with the values for other SLSNe-I from \cite{Nicholl2016b} and \cite{Chen2023b}, the latter of whom found a median velocity of 12\,800~km~s$^{-1}$ when studying the \ion{Fe}{II} lines in the ZTF SLSN-I sample. 

We note that the velocity measured from the \ion{Fe}{II} lines is higher than that measured from the tentative \ion{O}{II} lines, which is not unprecedented as shown in \cite{Quimby2018} and  \cite{Chen2023b}. However, the difference of $8000$~km~s$^{-1}$ has never been observed as the average difference between the estimated velocities using \ion{Fe}{II} and \ion{O}{II} is $\sim$ 3000~km~s$^{-1}$ \citep[e.g.,][]{Chen2023b}. A more likely explanation is that the absorption tentatively identified as \ion{O}{II} are dominated by different lines.

The low S/N in the cold photospheric phase spectra prevents us from tracking the evolution of the velocity. The strong feature at 4870~\AA\ in the spectrum at $+43.7$ days suggests that the \ion{Fe}{II} $\lambda5018$ may be blueshifted by $\sim$ 9000~km~s$^{-1}$, in which case the mismatch of the absorption at 4730~\AA\ can be explained by a blend of \ion{Fe}{II} with other ions. On the other hand, the absorption feature at 4740~\AA\ could be associated with the \ion{Fe}{II} $\lambda4923$ at $-12\,000$~km~s$^{-1}$, which would result in a constant velocity over a period of 40 days \citep{Nicholl2013,Nicholl2016a,Nicholl2016b,Liu2017}.

\section{Comparison to other C-rich SLSNe} \label{comparison}

In Sect.~\ref{light_curve_properties}, we compared the light curve properties of SN\,2020zbf with a homogeneous sample of SLSNe-I, concluding that the general photometric characteristics of SN\,2020zbf are overall average aside from the fast rise. However, the spectral properties, notably the strong \ion{C}{II} lines, are unusual in SLSNe though not entirely unprecedented. We inspected the available spectra in the literature and find a number of objects also noted for their strong \ion{C}{II} features. We note that the comparison is not with the full sample of all the C-rich objects but rather with a carefully selected sample of publicly available C-dominated SLSNe-I studied in individual papers as well as in sample papers. This includes SN\,2018bsz \citep{Anderson2018,Pursiainen2022}, SN\,2017gci \citep{Fiore2021}, SN\,2020wnt \citep{Gutirrez2022, Tinyanont2023}, iPTF16bad \citep{Yan2017} and PTF10aagc \citep{DeCia2018, Quimby2018}. In this section, we compare the properties of SN\,2020zbf to these other C-rich objects.

\subsection{Light curve comparisons} 

\label{photometry_comp}

\begin{figure}[h]
   \centering
   \includegraphics[width=0.5\textwidth]{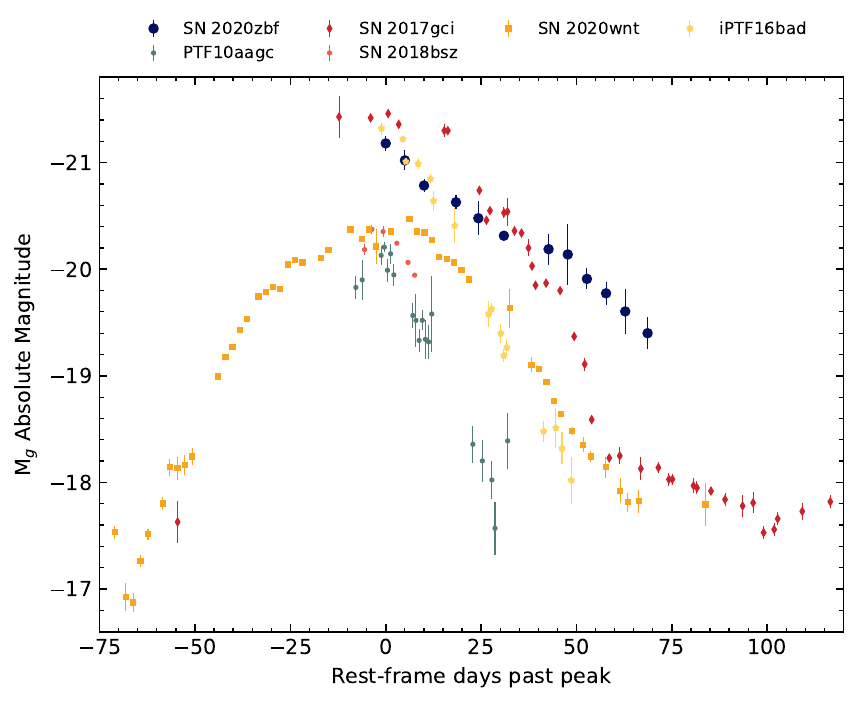}
   \caption{Rest-frame absolute magnitude $g$-band light curve of SN\,2020zbf in comparison with C-rich SLSNe-I from the literature. The magnitudes are K-corrected and corrected for MW extinction.}
              \label{LC_comparison}%
    \end{figure}

In Fig.~\ref{LC_comparison}, we compare the $g$-band light curve with $g$-band light curves of SLSNe-I that have been found to have strong \ion{C}{II} features in their early spectra. The absolute magnitudes of all the objects are corrected for the MW extinction and cosmological K-correction. C-rich SLSNe show a peak absolute magnitude distribution from $-20.1$ to $-21.5$ with a mean of $-20.8$, and SN\,2020zbf is in the upper half of the distribution. SN\,2020zbf fades by 1.8 mag in 79 days ($\sim $0.02 mag d$^{-1}$) and declines more slowly than the other SLSNe with \ion{C}{II} features. We notice a large diversity in the light curves of the C-rich sample, possibly reflecting the variety of mechanisms powering them and the general diversity among the whole population of SLSNe-I.

\subsection{Spectra comparisons}

\begin{figure*}
     \centering
     \includegraphics[width=\textwidth]{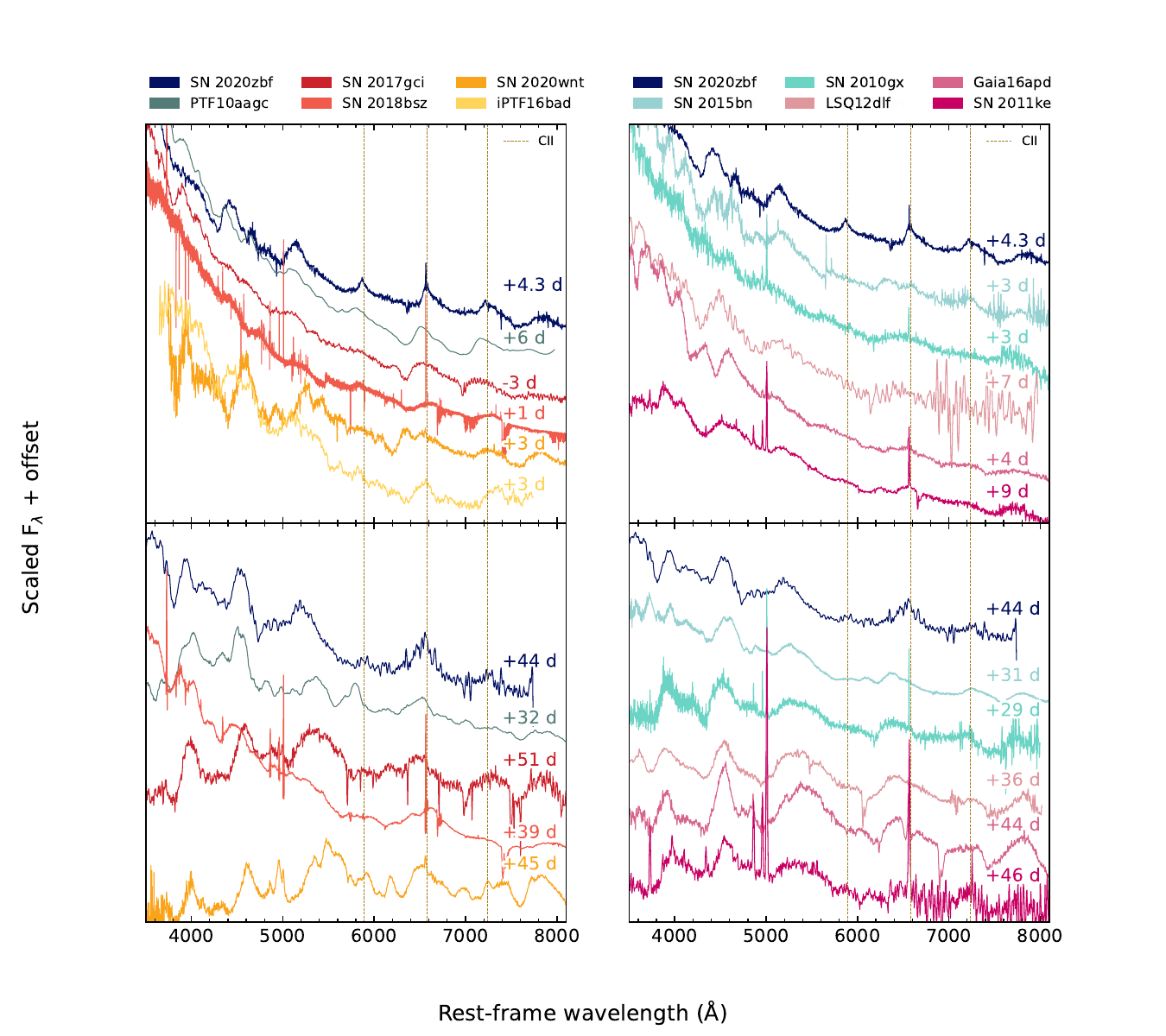}
\caption{Spectral comparison of SN\,2020zbf with SLSNe-I from the literature. Left: SN\,2020zbf spectra in comparison with C-rich SLSNe-I at around peak (top) and $\sim$ 30-50 days after peak (bottom). Right: Comparison of SN\,2020zbf spectra with typical well-studied SLSNe-I at the same epochs. The spectra are corrected for MW extinction. The spectra of SN\,2020zbf, SN\,2018bsz, iPTF16bad and LSQ12dlf have been smoothed using a Savitzky-Golay filter. The vertical gold dashed lines indicate the rest wavelengths of the \ion{C}{II} emission lines.}

\label{spectra_comp}%
\end{figure*}

In Fig.~\ref{spectra_comp} (left panel), we compare the spectra of SN\,2020zbf with those of the C-rich sample. We used two reference epochs: around peak (top) and $\sim$ $+40$ days post-maximum light (bottom). As mentioned above, the spectrum of SN\,2020zbf at $+4.3$ days presents strong \ion{C}{II} $\lambda 5890$, $\lambda 6580$, $\lambda 7234$, and \ion{Fe}{III} $\lambda 4432$ and $\lambda 5129$. These characteristics match well with the near peak spectrum of PTF10aagc \citep{Quimby2018} but are stronger in SN\,2020zbf. Even though the \ion{C}{II} profiles of PTF10aagc show similarities, the peak of the lines is blueshifted by a higher velocity (3000~km~s$^{-1}$) for that SN than for SN\,2020zbf. The \ion{C}{II} lines in PTF10aagc also show an absorption component in the blue; in SN\,2020zbf absorption components are not obviously visible. Strong \ion{C}{II} lines are also observed in SN\,2017gci \citep{Fiore2021} though the shapes are different compared to SN\,2020zbf and the 5890~\AA\ emission is absent. The \ion{C}{II} $\lambda6580$ and $\lambda7234$ in SN\,2017gci are broader than in SN\,2020zbf and P-Cygni profiles are present. SN\,2018bsz is a SLSN-I that has been studied particularly for the strong \ion{C}{II} features it shows in its spectra; however, the \ion{C}{II} features in SN 2018bsz are weaker than those of SN\,2020zbf, and the general structure of the spectrum is different. SN\,2020wnt \citep{Gutirrez2022,Tinyanont2023} and iPTF16bad \citep{Yan2017} also present \ion{C}{II} $\lambda6580$ and $\lambda7234$ lines, yet their spectra are significantly distinct from those of the other objects in the sample. SN\,2020wnt is the only object of the sample that shows \ion{Si}{II} at 6300~\AA\ and a strong feature at 5200~\AA\ mostly attributed to \ion{Fe}{II} (see Sect.~\ref{photometry_comp}).

A large variety regarding the presence of the \ion{O}{II} features in the spectra exist in the C-rich sample. In SN\,2020zbf we can distinguish only the \ion{O}{II} $\lambda4358$ and $\lambda4651$ W-shape, which appears to match well with the \ion{O}{II} lines of PTF10aagc. For SN\,2017gci, even though it exhibits these features, they are weaker than in SN\,2020zbf. In contrast, SN\,2018bsz seems to not show any features at these wavelengths at the considered epoch, and \cite{Gutirrez2022} and \cite{Tinyanont2023} demonstrate that the spectrum of SN\,2020wnt lacks the \ion{O}{II} feature. In the case of iPTF16bad it is unclear whether it presents \ion{O}{II} since only the feature at 4651~\AA\ is visible. We conclude that apart from PTF10aagc, which is a relatively good match, none of the C-rich SLSNe-I in this sample resembles SN\,2020zbf in terms of the shape and the intensity of the \ion{C}{II} or \ion{O}{II} ion lines or resembles some other object of the sample. This diversity in the spectra around peak is in concordance with the variety of the light curve shapes we found in Sect.~\ref{photometry_comp}.

In Fig.~\ref{spectra_comp} (right panel), we compare SN\,2020zbf with a sample of well-studied SLSNe-I including SN\,2015bn \citep{Nicholl2016a}, SN\,2010gx \citep{Pastorello2010}, LSQ12dlf \citep{Nicholl2014}, Gaia16apd \citep{Kangas2017}, and SN\,2011ke \citep{Inserra2013}. Typically, the red part of the spectrum of SLSNe-I shows weak features of \ion{O}{II} and \ion{C}{II} \citep{GalYam2019b}, which we observe in this sample. However, the \ion{C}{II} lines of SN\,2020zbf have a unique shape and strength that set it apart from the other objects with a similar general spectral shape. In addition, the \ion{O}{II} W-shape appears to be present in these SLSNe-I though it is typically shifted by different velocities. SN\,2011ke is an exception since these features are not present in the spectra at this particular epoch.

SN\,2020zbf has similar properties in the late-time spectra as typical SLSNe-I at these phases and also to SNe-Ic at around peak \citep{GalYam2019b}. In contrast to SN\,2020zbf, which exhibits no \ion{Si}{II} $\lambda6355$, all the comparison objects show a strong \ion{Si}{II} emission line, but none of the events presents the strong feature at around $\sim$ 6580~\AA, which is persistent in all the C-rich objects and evolves to H$\alpha$ in the majority of them \citep{Yan2015,Yan2017,Fiore2021,Pursiainen2022}.

\section{Comparison to model spectra} \label{sec:synthetic_spectra}

\begin{figure}
     \centering
     \includegraphics[width=0.5\textwidth]{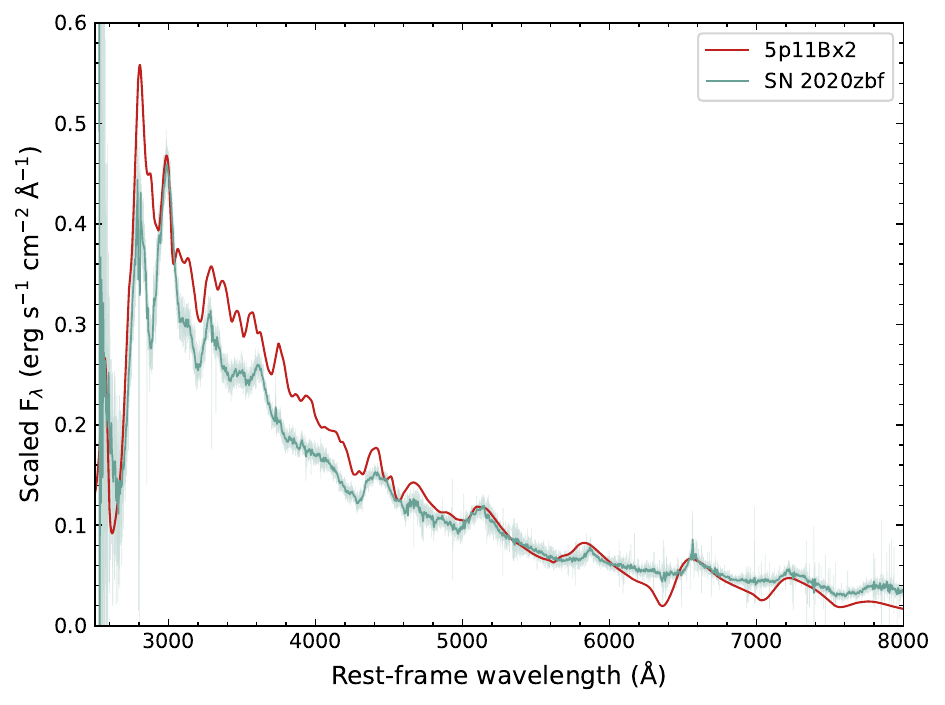}
\caption{Comparison of the high-quality X-shooter spectrum of SN\,2020zbf at $\sim$ $+4.3$ days post maximum (green) with a synthetic spectrum at 12 days after peak from \cite{Dessart2019} (red). The spectra are corrected for MW extinction and
the X-shooter spectrum is smoothed using a Savitzky-Golay filter. The original spectrum is shown in lighter colors, whereas the telluric absorptions are
denoted by lighter gray vertical bands. }
\label{model_spectra}%
\end{figure}

We compared the X-shooter spectrum of SN\,2020zbf at $+4.3$ days after maximum with synthetic model spectra presented in \cite{Dessart2019}. These spectra are the result of time-dependent radiative transfer simulations based on magnetar-powered SNe-Ic using the numerical approach of \cite{Dessart2018}. The SNe are followed with the nonlocal thermodynamic equilibrium radiative transfer code \program{CMFGEN} \citep{Hillier2012} from day one until one or two years after the explosion \citep{Dessart2015,Dessart2016,Dessart2017a}.

Exploring the whole range of models presented in \cite{Dessart2019}, we find that the best match to the observed data is the model 5p11Bx2. The progenitor of this model is 5p11, which is described in \cite{Yoon2010} and corresponds to a 60~$\rm M_\odot$ ZAMS star with solar metallicity; it is a primary star of a close binary system (orbital period of 7 days) that evolves to a WR star with a final mass of 5.11~$\rm M_\odot$ (rather than the 4.95~$\rm M_\odot$ quoted in \citealt{Yoon2010}; see \citealt{Dessart2015}). This model has lost its He-rich envelope and the surface C mass fraction $X_{\rm C,s}$ is computed to be 0.51 \citep{Dessart2016}. As suggested by \cite{Yoon2010}, the end fate of this progenitor is more likely to be a SN-Ic.

The explosion is then remapped with \program{v1D} \citep{Livne1993,Dessart2010a,Dessart2010b} by means of a piston producing a kinetic energy of 2.49~$\times$~10$^{51}$~erg (suffix B in 5p11Bx2), which accounts for the additional kinetic energy to match SLSN ejecta velocities. This model also uses a strong mixing process in the ejecta (suffix x2 in 5p11Bx2; \citealt{Dessart2015}). According to \cite{Dessart2016}, the 5p11Bx2 model corresponds to a H-deficient ejecta of 3.63~$\rm M_\odot$ expanding with 11\,500~km~s$^{-1}$ that contains 0.34~$\rm M_\odot$ of He, 0.89~$\rm M_\odot$ of C, 1.40~$\rm M_\odot$ of O, 0.15~$\rm M_\odot$ of Si, and  0.19~$\rm M_\odot$ of $^{56}$Ni (mass prior to the decay). The magnetar in this model has an initial rotational energy of 0.4~$\times$~10$^{51}$~erg, a magnetic field of 3.5~$\times$~10$^{14}$~G, an initial spin period of 7~ms and a spin-down timescale of 19.1 days.

In Fig.~\ref{model_spectra}, the observed X-shooter spectrum is compared to the synthetic one. The best match for the observed spectrum ($+4.3$ days past maximum) is obtained for a model at 38.4 days after explosion corresponding to $+11.7$ days after maximum. We highlight that the model spectrum has not been developed or fine-tuned to match SN\,2020zbf, but rather employs the previously described grid of parameters. However, it qualitatively reproduces the overall shape and the main features of the observed spectrum, despite the fact that the model lines are slightly blueshifted and the temperature of the model is higher. In the red part of the spectrum, the model matches almost perfectly the profiles associated with \ion{Fe}{II} $\lambda5169$ and \ion{Fe}{III} $\lambda5129$, \ion{C}{II} $\lambda5890$, $\lambda6580$, $\lambda7234,$ and \ion{O}{I} $\lambda\lambda7772, 7774, 7775$. In the synthetic spectra, the \ion{C}{II} form P-Cygni profiles that we do not observe in the spectra of SN\,2020zbf (see the discussion in Sect.~\ref{sec:cii}). Although weak \ion{He}{I} $\lambda5875$ is present in the model, the line is mostly dominated by \ion{C}{II}, which is in agreement with our identification scheme. The computed spectrum in the blue part shows contributions of \ion{Mg}{II}, \ion{C}{II,} and \ion{C}{III} at 2670~\AA, \ion{Mg}{II} at 2880~\AA, \ion{Fe}{III} in the region 3000 -- 3600~\AA, \ion{O}{II} between 3500 and 5000~\AA,\ and \ion{Mg}{II} at 4330~\AA. 

The 5p11Bx2 model predicts a rise time of 26.7 days and a maximum bolometric luminosity of 4.86~$\times$~10$^{43}$~erg~s$^{-1}$ (see Table~1 in \citealt{Dessart2019}). In the photometric analysis of SN\,2020zbf (see Sect.~\ref{light_curve_properties} and \ref{bolometric_properties}), we estimated the rise time to be $ \lesssim 26.4$ rest-frame days and the peak bolometric luminosity to be $\gtrsim$ $8.34$ $\pm$ $0.49$~$\times$~10$^{43}$~erg~s$^{-1}$. Since 5p11Bx2 has not been modeled to match SN\,2020zbf, we expect some differences in the light curve properties. However, both the observed upper limit in the rise time and the modeled value are placed in the fastest regime of the SLSNe-I. On the other hand, the estimated peak bolometric luminosity of SN\,2020zbf is 1.7~times higher than that of the model. Other works that find good matches to \cite{Dessart2019} spectral models also find discrepancies in the light curve behaviors. For example, \cite{Anderson2018} find a good match between spectral models and the observed spectra of SN\,2018bsz but the light curve models did not agree with the observations. They interpreted this fact as due to the model missing a mechanism that could lead to the slow rise of SN\,2018bsz. \cite{Anderson2018} also point out that the input kinetic energy of the model and/or the magnetar energy deposition profile can affect both the rise time and the maximum bolometric luminosity.

In conclusion, the 5p11Bx2 model produces a synthetic spectrum that matches the general shape of the X-shooter data very well  and reproduces the majority of the features we observe in SN\,2020zbf. However, the selection of other parameters such as the mass of the progenitor, the ejecta mass, the amount of C in the ejecta and the kinetic energy might lead to a light curve and spectral properties closer to the observed ones. We explore the magnetar models for the light curve of SN\,2020zbf in Sect.~\ref{sec:magnetar}.

\section{Light curve modeling} \label{sec:lc_model}

In this section, we model the observed multiband light curves of SN\,2020zbf using the Python-based Modular Open Source Fitter for Transients (MOSFiT) code (\citealt{Guillochon2018}) to determine the most likely powering source. Giving as input the SN redshift, the luminosity distance, the $E(B-V)$ from the MW, the light curve data in the different bands and a set of priors listed in Table~\ref{tab:mosfit_results}, MOSFiT outputs the posterior distribution of the modeled light curve parameters. The prior for the ejecta velocity was set to be a Gaussian distribution with $\mu$ = 10\,000~km~s$^{-1}$ and $\sigma$ = 2000~km~s$^{-1}$.

We selected the \program{default} $^{56}$Ni model based on the paper of \cite{Arnett1982} and \cite{Nadyozhin1994}, the \program{slsn} magnetar model described in \cite{Nicholl2017} and the \program{csm} model based on the semi-analytic model from \cite{Chatzopoulos2013}. These models are fitted using the dynamic nested sampling package \program{dynesty} \citep{Speagle2020}. The quality of the fit can be quantified by the likelihood score (logZ; \citealt{Watanabe2010}) for each model. 

\subsection{Radioactive source model}

We tested the hypothesis that the light curve of SN\,2020zbf is powered by the radioactive decay of $^{56}$Ni \citep{Arnett1982}. The resulting fits to the observed data are plotted in Fig.~\ref{fig:mosfit_ni} and the values of the parameters are listed in Table~\ref{tab:mosfit_results}. The corner plot is presented in Fig.~\ref{fig:corner_ni}.

By a visual inspection, the model does not reproduce the multiband light curves very well. It fits the rising part in the $c$ filter but it does not capture the observed peak magnitudes in the $g$ and $B$ bands and it does not fit the UVOT data. In addition, the model light curves decline more slowly in the bluer bands than the observed ones, and at $\sim$ +180 days the pure $^{56}$Ni model fails to describe the data.

According to the results from MOSFiT, the high luminosity of SN\,2020zbf requires an unrealistic amount of 90\% $^{56}$Ni in the 7.9~$\rm M_\odot$ ejecta. We conclude that radioactive decay is unlikely to be the main contributor to the luminosity for SN\,2020zbf, which is consistent with the conclusions for other H-poor SLSNe \citep[e.g.,][]{Chomiuk2011,Quimby2011,Inserra2013,Nicholl2017,Chen2023b}.

\begin{figure*}
     \centering
     \begin{subfigure}[b]{0.9\textwidth}
         \centering
         \includegraphics[width=\textwidth]{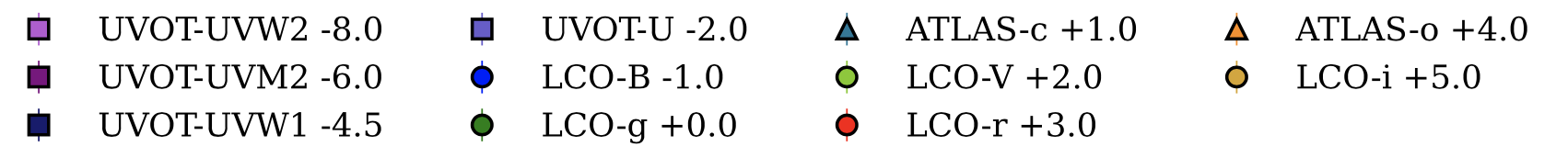}
         %\caption{}
         \label{fig:mosfit_ni}
     \end{subfigure}
     \begin{subfigure}[b]{0.33\textwidth}
         \centering
         \includegraphics[width=\textwidth]{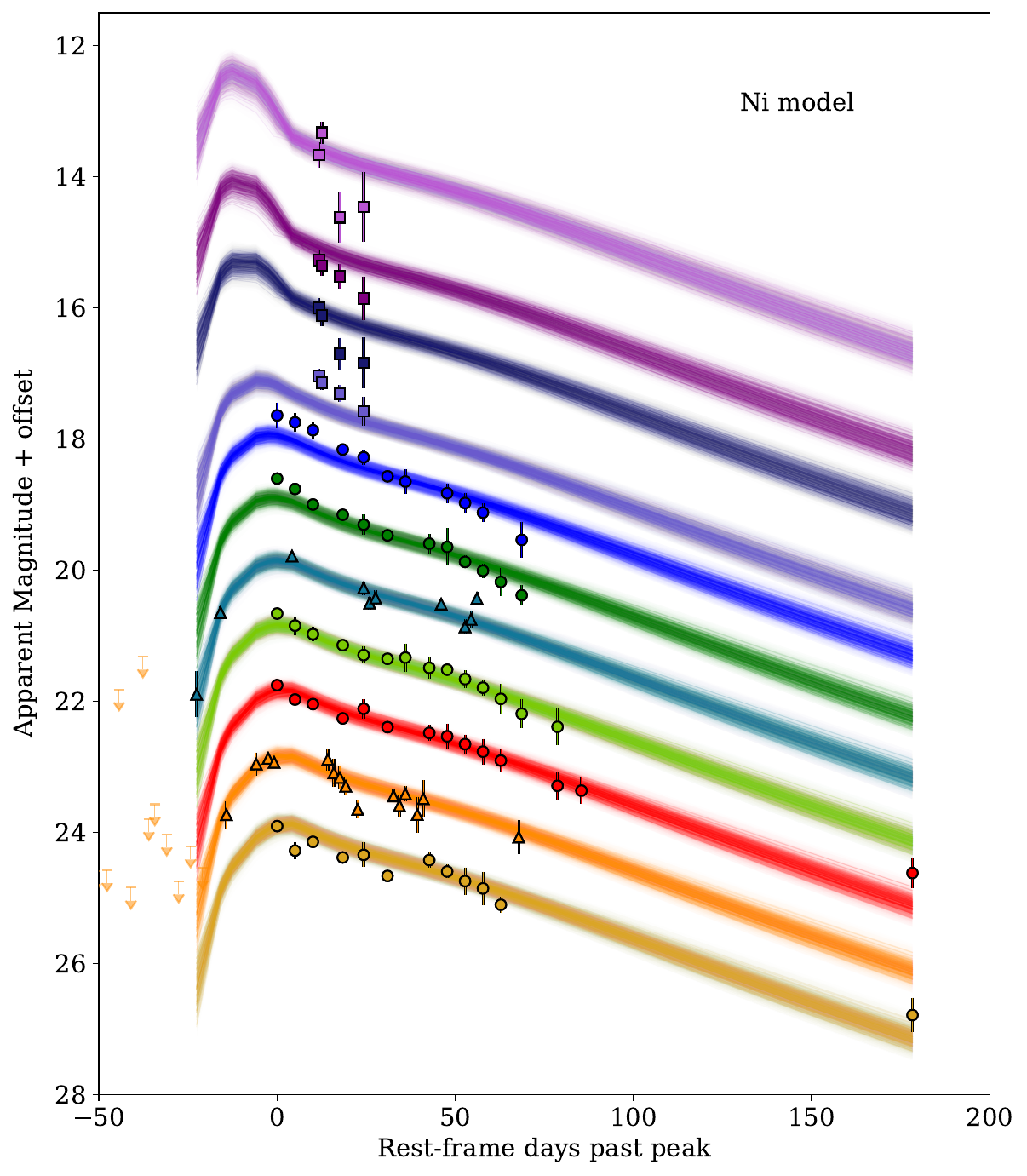}
         \caption{}
         \label{fig:mosfit_ni}
     \end{subfigure}
     %\hfill
     \begin{subfigure}[b]{0.33\textwidth}
         \centering
         \includegraphics[width=\textwidth]{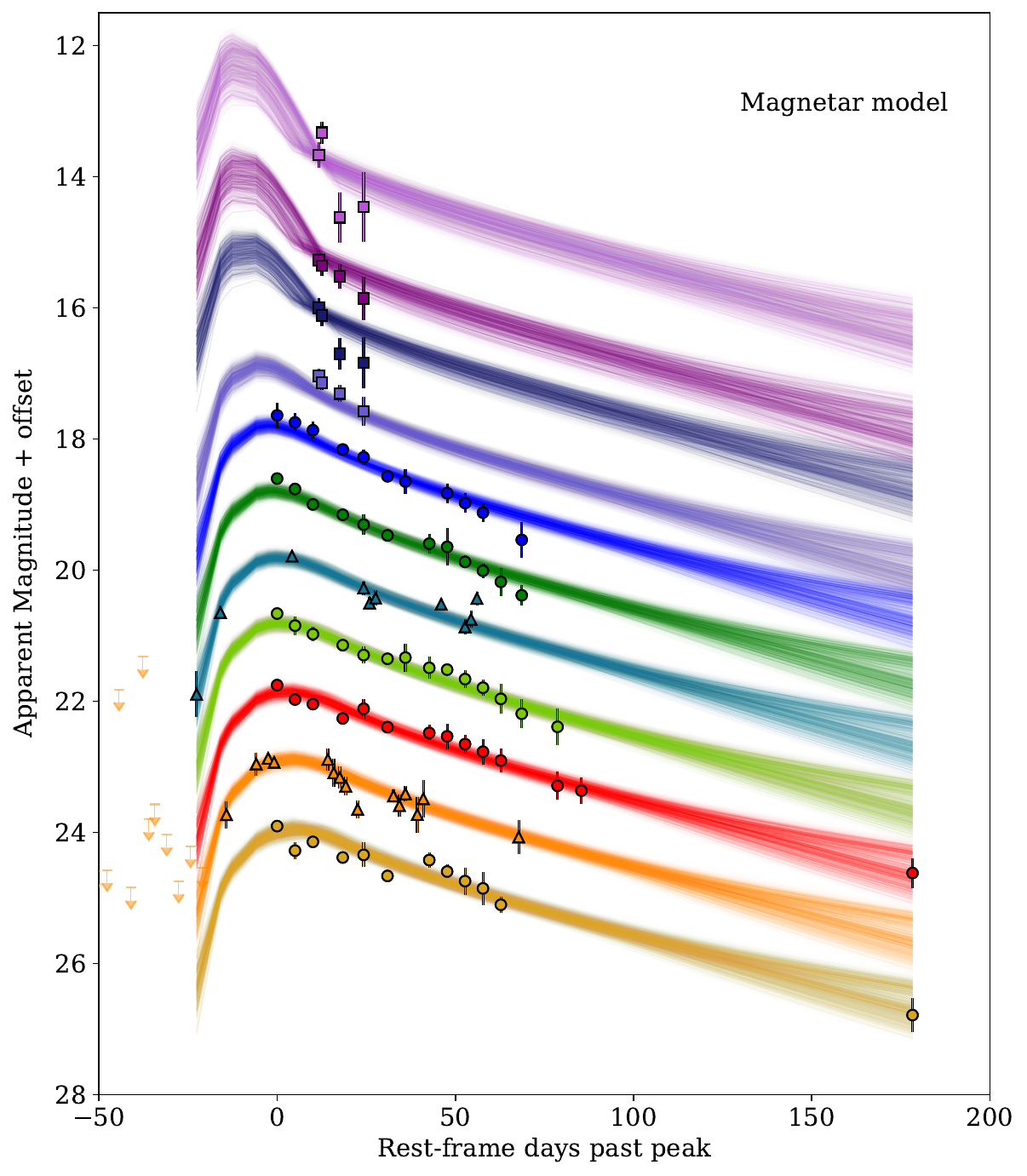}
         \caption{}
         \label{fig:mosfit_slsn}
     \end{subfigure}
     \begin{subfigure}[b]{0.33\textwidth}
         \centering
         \includegraphics[width=\textwidth]{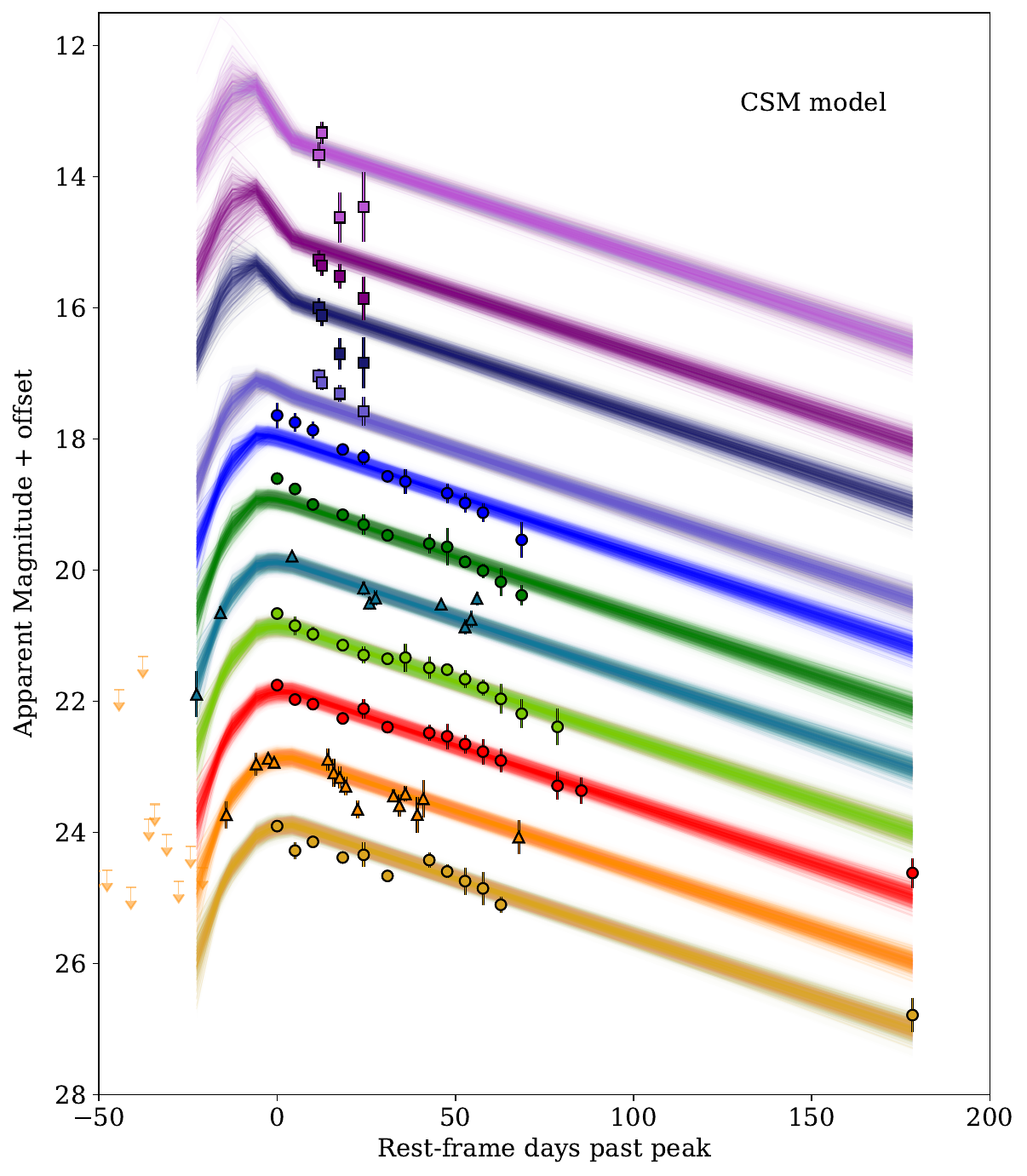}
         \caption{}
         \label{fig:mosfit_csm}
     \end{subfigure}
\caption{Multiband light curves of SN\,2020zbf and their fits with the three different models in MOSFiT: Ni model (panel a), the magnetar model (b), and the \program{csm} model (c). The colored lines indicate the range of the most likely fits.}
\end{figure*}

\begin{table}
\centering
\caption{Priors and posterior of the parameters fitted with MOSFiT for the \program{default} , \program{slsn,} and \program{csm} model. The $\mathcal{U}$ stands for uniform, $\mathcal{L}$ for log-uniform, and $\mathcal{G}$ for Gaussian.}
\label{tab:mosfit_results}
\begin{tabular}{lll}

\hline
\hline

Parameters & Priors & Best-fit values  \\
                
\hline

&  $\rm ^{56}Ni$ model & \\ 
\hline

$f_{\rm ^{56}Ni}$ & $\mathcal{U}$ (10$^{-3}$, 1) & 0.9$^{+0.05}_{-0.1}$ \\
$M_{\rm ej}$ ($\rm M_\odot$) & $\mathcal{L}$ (0.1, 100)  & 7.9$^{+1}_{-0.5}$ \\
$\rm \kappa$ (cm$^{2}$ g$^{-1}$) & $\mathcal{U}$ (0.05, 0.2) & 0.06$^{+0.01}$ \\
$\rm \kappa_\gamma$ (cm$^{2}$ g$^{-1}$) & $\mathcal{L}$ ($10^{-4}$, $10^4$) & 0.05$^{+0.02}_{-0.01}$ \\
$T_{\rm min}$ (K) & $\mathcal{U}$ (3~$\times$~10$^{3}$, 2~$\times$~10$^{4}$) & 10\,178$^{+169}_{-162}$\\
$v_{\rm ej}$ (km s$^{-1}$) & $\mathcal{G}$ ($\mu = 10^{4}$, $\sigma = 2~\times$~10$^{3}$) & 13\,535$^{+918}_{-875}$ \\
Score (log Z)  & & 86.12 \\
\hline

&  Magnetar-powered model & \\ 
\hline

$M_{\rm ej}$ ($\rm M_\odot$) & $\mathcal{L}$ (0.1, 100) & 1.5$^{+0.6}_{-0.4}$ \\
$M_{\rm NS}$ ($\rm M_\odot$) &$\mathcal{U}$ (1, 2) & 1.6$^{+0.2}_{-0.2}$\\
$B$ (10$^{14}$ G) & $\mathcal{U}$ (0.1, 10) & 2.2$^{+3.4}_{-1}$ \\
$P_{\rm spin}$ (ms) & $\mathcal{U}$ (1, 10) & 5.1$^{+0.5}_{-0.7}$ \\
$\rm \kappa$ (cm$^{2}$ g$^{-1}$) & $\mathcal{U}$ (0.05, 0.2) & 0.12$^{+0.04}_{-0.04}$ \\
$\rm \kappa_\gamma$ (cm$^{2}$ g$^{-1}$) & $\mathcal{L}$ ($10^{-4}$, $10^4$) & 0.7$^{+35.6}_{-0.3}$ \\
$T_{\rm min}$ (K) & $\mathcal{U}$ (3~$\times$~10$^{3}$, 2~$\times$~10$^{4}$) & 10\,844$^{+271}_{-387}$ \\
$v_{\rm ej}$ (km~s$^{-1}$) & $\mathcal{G}$ ($\mu = 10^{4}$, $ \sigma = 2~\times~10^{3}$) & 9603$^{+1304}_{-952}$ \\
Score (log Z) & & 129.29 \\

\hline

&  CSM  model & \\ 
\hline
$M_{\rm ej}$ ($\rm M_\odot$) & $\mathcal{L}$ (0.1, 100) & 0.2$^{+0.1}_{-0.1}$ \\
$M_{\rm CSM}$ ($\rm M_\odot$) & $\mathcal{L}$ (0.1, 30) & 4.8$^{+0.7}_{-0.5}$ \\
$\rho$ ($10^{-12}$~g~cm$^{-3}$) & $\mathcal{L}$ ($10^{-15}$, $10^{-11}$) & 1.15$^{+0.66}_{-0.34}$ \\
$s$  & $\mathcal{U}$ (0, 2) & 0.21$^{+0.21}_{-0.13}$ \\
$T_{\rm min}$ (K) & $\mathcal{U}$ (3~$\times$~10$^{3}$, 2~$\times$~10$^{4}$) & 10\,137$^{+122}_{-121}$ \\
$v_{\rm ej}$ (km s$^{-1}$) & $\mathcal{G}$ ($\mu = 10^{4}$, $\sigma = 2~\times~10^{3}$) & 12\,343$^{+948}_{-1068}$ \\
Score (log Z) & & 115.97 \\
\hline

\hline
\end{tabular}
\end{table}

\subsection{Magnetar source model} \label{sec:magnetar}

Assuming that SN\,2020zbf is powered by the spin-down of a rapidly rotating newly formed neutron star \citep{Ostriker1971,Arnett1989,Kasen2010,Chatzopoulos2012,Inserra2013}, we ran the \program{slsn} model in MOSFiT setting similar priors to those used by \cite{Nicholl2017}. This model assumes a modified SED accounting for the line blanketing in the UV part of the SLSN spectra \citep{Chomiuk2011}. The MOSFiT light curve fits are shown in Fig.~\ref{fig:mosfit_slsn} and the resulting values of the posteriors in Table~\ref{tab:mosfit_results}. The corner plot is presented in Fig.~\ref{fig:corner_slsn} (see Appendix~\ref{app:mosfit}). Visually, the model captures the rise, peak and decline up to $\sim$ $+50$ days and it fits the data point at $\sim$ $+ 180$ days. However, after $\sim+50$ days, the light curves in the bluer bands decline faster than the model. 

The inferred spin period of 5.14~ms is higher compared to the typical values ($\sim$ 2.5~ms) found in the literature, approaching the slowest limit of the observed SLSNe thought to be powered by a magnetar \citep{Nicholl2017,Chen2023b,Blanchard2020}. Similarly, the value of the magnetic field (2.18~$\times$~10$^{14}$~G) falls in the highest regime but within the SLSN magnetar properties \citep{Nicholl2017,Chen2023b}. The resulting ejecta mass from the model (1.51~$\rm M_\odot$) is significantly lower than the results of \cite{Chen2023b}, who found that the median ejecta mass is $5.03^{+4.01}_{-2.39}$~$\rm M_\odot$. Given that SN\,2020zbf is placed among the fastest rising SLSNe in the \cite{Chen2023a} sample, we expect the ejecta mass of SN\,2020zbf to be significantly lower than the median value of the ZTF sample. This is also lower than the findings of \cite{Nicholl2017}, who found a median of $4.8^{+8.1}_{-2.6}$~$\rm M_\odot$ studying a more heterogeneous sample of SLSNe-I. However, in the sample \citep{Chen2023b}, five SLSNe have ejecta masses of $< $2~$\rm M_\odot$. In addition, there are two further SLSNe with $M_{\rm ej}$ $<$ 2~$\rm M_\odot$: PS1-10bzj ($M_{\rm ej}$ = 1.65~$\rm M_\odot$; \citealt{Lunnan2013,Nicholl2017}) and SNLS-07D2bv ($M_{\rm ej}$ = 1.55~$\rm M_\odot$; \citealt{Howell2013}). Consequently, even though the predicted ejecta mass for SN\,2020zbf is lower than the median value, there are a few magnetar-powered SLSNe-I with similar derived ejecta masses. 

Based on the results of the model, we estimate the kinetic energy of the ejecta [$ E_{\rm k} = 0.6 \times 10^{51} \, (M_{\rm ej}/1 \, \rm M_\odot) \, (v_{\rm ej}/10^{4} \, \rm km \, \rm s^{-1})^{2}$~erg, uniform density] to be $0.84 \times 10^{51}$ erg, the rotational energy of the magnetar [$E_{\rm rot} = 2 \times 10^{52}P_{\rm ms}^{-2}(M_{\rm NS}/1.4 \, \rm M_\odot)^{3/2}$ erg; \citealt{Kasen2017}] to be $1.17 \times  10^{51}$~erg, the magnetar spin-down timescale [$\tau_{\rm spin} = 1.3 \times 10^{5}(M_{\rm NS}/1.4\, \rm M_\odot)^{3/2}P_{\rm ms}^{2}B_{\rm 14}^{-2}$ s; \citealt{Nicholl2017}] to be 10.4 days and the diffusion time [$\tau_{\rm diff} = 8.6 \times 10^{5} \, (M_{\rm ej}/1 \, \rm M_\odot)^{3/4} \, (\kappa/0.1 \, \rm cm^{2} \, \rm g^{-1})^{1/2} \, (E_{\rm k}/10^{51} \, \rm erg)^{-1/4}$~s; \citealt{Arnett1982}] to be 17.4 days. The ratio of the two timescales is $\sim$ 1.7, which according to \cite{Suzuki2021} indicates that a large fraction of the rotational energy is contributing to the SN luminosity, typical for SLSNe. This ratio is similar to the values found in \cite{Chen2023a,Chen2023b}. The estimated kinetic energy is lower than the median value of 2.13~$\times$~10$^{51}$~erg presented in \cite{Chen2023b}, which is a result of the lower ejecta mass. We estimate the total radiated energy of the SN [$E_{\rm k} = 10^{51} +  1/2(E_{\rm rot}-E_{\rm rad})$~erg; \citealt{Inserra2013}] to be 1.5~$\times$~10$^{51}$~erg, higher than our estimated integrated radiated energy of  0.47 $\pm$ 0.014~$\times$~10$^{51}$~erg. This discrepancy is mostly attributed to the energy emitted by the model in the UV around peak, which is not captured in our estimate since it is before our earliest UVOT data point.

We compared the results of MOSFiT with the magnetar parameters of the 5p11Bx2 model discussed in Sect.~\ref{sec:synthetic_spectra}. The model refers to a slow rotating magnetar (7~ms) with strong magnetic field (3.5~$\times$~10$^{14}$~G), properties that resemble the MOSFiT results but are placed in the most extreme limits of the \cite{Chen2023b} magnetar properties. The $M_{\rm ej}$ is 3.63~$\rm M_\odot$ in the 5p11Bx2 model instead of 1.51~$\rm M_\odot$ from MOSFiT. The overall good match of the magnetar models to the observed spectra and light curve allows us to favor a low ejecta mass explosion powered by a magnetar engine.

\subsection{CSM source model}

We modeled the multiband light curve of SN\,2020zbf using the \program{csm} model \citep{Chatzopoulos2013} in MOSFiT under the assumption that the light curve is dominated by CSM interaction. The light curve fits are depicted in Fig.~\ref{fig:mosfit_csm} and the resulting posteriors are listed in Table~\ref{tab:mosfit_results}. The corner plot is presented in Fig.~\ref{fig:corner_csm} (see Appendix~\ref{app:mosfit}). The model captures the rise but not the peak in the $B$ and $g$ band, declines more slowly than the data and fails to fit the single data point at $\sim$ $+180$ days.

The resulting CSM mass is estimated to be 4.79~$\rm M_\odot$, and the ejecta mass is 0.22 $\rm M_\odot$, which is even lower than the ejecta mass inferred by the \program{slsn} model. In \cite{Chen2023b} the median value of the CSM mass is $4.67^{+6.90}_{-2.56}$~$\rm M_\odot$, similar to our results (4.79~$\rm M_\odot$), whereas the ejecta mass derived from the CSM model for their sample is 11.92$^{+24.98}_{-10.65}$ $\rm M_\odot$. The wide distribution of the mass reflects the relatively large amount of events with very low $M_{\rm ej}$. In particular, 11 out of 70 SLSNe-I have ejecta masses similar to SN\,2020zbf and the corresponding CSM masses span the whole range of the distribution. Three out of these 11 objects favor the magnetar over the CSM model, six events can be fit equally well by both the magnetar and CSM models and two clearly prefer the CSM fit. Hence, there are SLSNe-I that can be well fit by an CSM model with very low ejecta masses (< 1~$\rm M_\odot$).

It has been demonstrated that using the same CSM structure, the semi-analytic model from \cite{Chatzopoulos2013} and hydrodynamic simulations can generate conflicting results \citep{Moriya2013,Sorokina2016,Moriya2018} and the quantitative values of the CSM parameters may only be an order of magnitude approximation. Additionally, it is not clear what kind of progenitor scenario would result in such a configuration, with an extremely stripped star (0.2~$\rm M_\odot$ ejecta) exploding into massive ($> 4~$M$_\odot$) carbon-oxygen CSM. We conclude that the light curve shape is also consistent with a CSM model, but more careful modeling outside the scope of this paper is necessary to explore the possible progenitor and CSM structure.

\section{Host galaxy} \label{sec:host}

The left panel of Fig.~\ref{fig:2020zbf_image} shows the Legacy Survey image of the field around SN\,2020zbf. A small galaxy is visible at the SN location; the reported LS photometry corresponds to an absolute magnitude $M_{\rm g} = -17.1$~mag (at an effective wavelength of $\sim$ 4000~\AA), similar to the Large Magellanic Cloud. 

\subsection{Galaxy SED modeling}

\begin{figure}
    \centering
    \includegraphics[width=1\columnwidth]{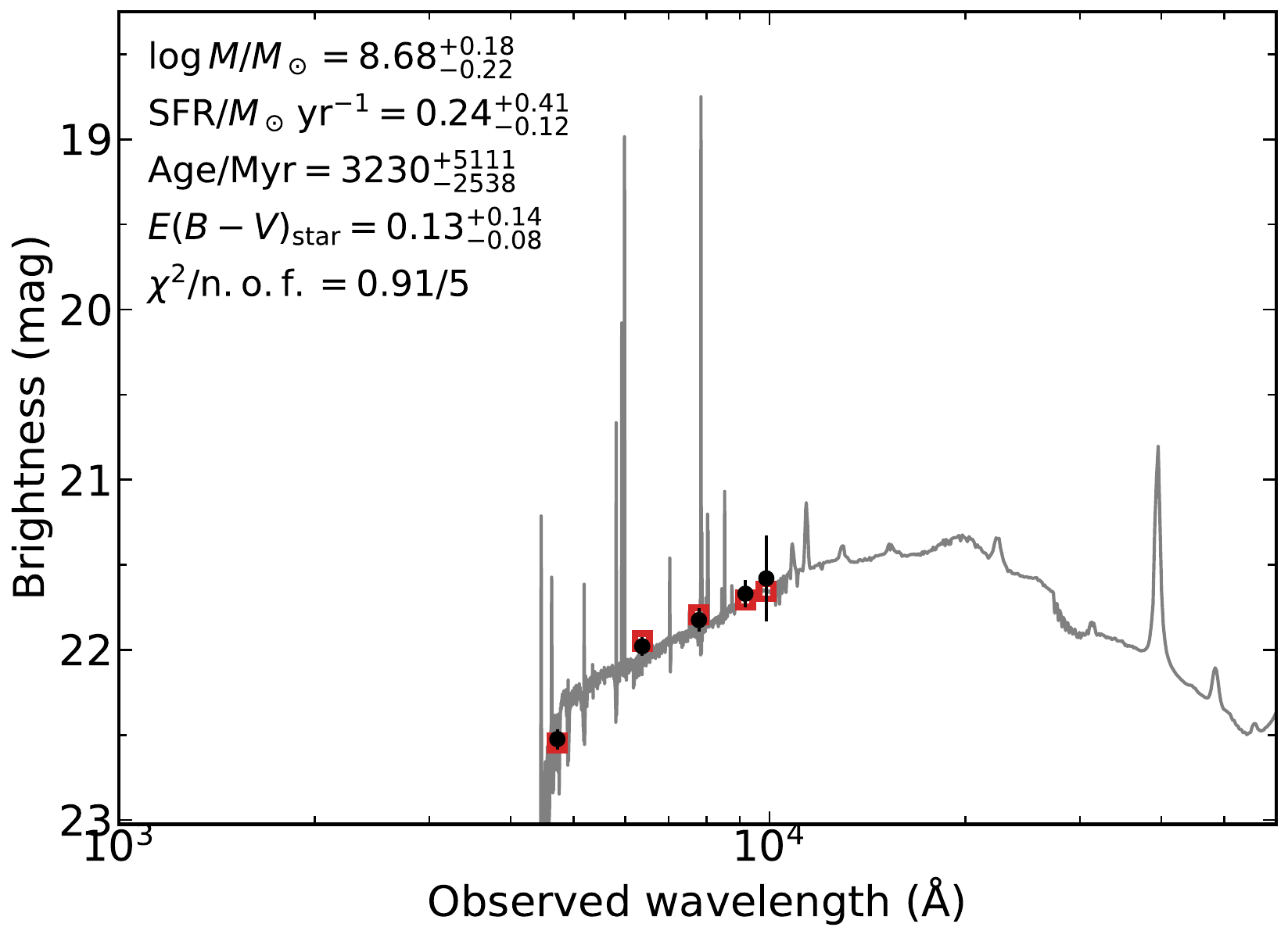}
    \caption{SED of the host galaxy from 1000 to 60\,000~\AA\ (black data points). The solid line represents the best-fitting model of the SED. The red squares represent the model-predicted magnitudes. The fitting parameters are shown in the upper-left corner. The abbreviation ``n.o.f.'' stands for the number of filters.}
    \label{fig:host_sed}
\end{figure}

We modeled the observed SED (black data points in Fig.~\ref{fig:host_sed}, tabulated in Table~\ref{tab:host_phot}) with the software package \program{Prospector} version 1.1 \citep{Johnson2021a}.\footnote{\program{Prospector} uses the \program{Flexible Stellar Population Synthesis} (\program{FSPS}) code \citep{Conroy2009a} to generate the underlying physical model and \program{python-fsps} \citep{ForemanMackey2014a} to interface with \program{FSPS} in \program{python}. The \program{FSPS} code also accounts for the contribution from the diffuse gas based on the \program{Cloudy} models from \citet{Byler2017a}. We used the dynamic nested sampling package \program{dynesty} \citep{Speagle2020} to sample the posterior probability.} We assumed a Chabrier initial mass function \citep{Chabrier2003a} and approximated the star formation history (SFH) as a linearly increasing SFH at early times followed by an exponential decline at late times [functional form $t \times \exp\left(-t/\tau\right)$, where $t$ is the age of the SFH episode and $\tau$ is the $e$-folding timescale]. The model is attenuated with the \citet{Calzetti2000a} model. The priors of the model parameters are set identically to those used by \citet{Schulze2021a}. 

Figure \ref{fig:host_sed} shows the observed SED (black dots) and its best fit (gray curve). The SED is adequately described by a galaxy template with a log$_{10}$ mass of $8.68^{+0.18}_{-0.22}~\rm M_\odot$, and a star formation rate of $0.24^{+0.41}_{-0.12}~\rm M_\odot\,{\rm yr}^{-1}$.

\subsection{Emission line diagnostics}

The X-shooter spectrum shows a number of narrow emission lines from the galaxy seen atop the SN continuum (see Fig.~\ref{fig:hostlines}). After calibrating this spectrum to the photometry and correcting for MW extinction, we measured the line fluxes by fitting Gaussian line profiles. The resulting flux values are listed in Table~\ref{tab:line_fluxes}. 

\begin{table}
\centering
\small
\caption{ Observed host galaxy emission line fluxes (corrected for MW extinction).}
\label{tab:line_fluxes}
\begin{tabular}{cc}
\hline
\hline
Line         & Flux       \\
           &  ($10^{-17}~{\rm erg~s}^{-1}{\rm cm}^{-2}$) \\ \hline
$[$\ion{S}{II}$]~\lambda6731$ &  $1.61 \pm 0.28$ \\
$[$\ion{S}{II}$]~\lambda6717$ &  $2.27 \pm 0.37$ \\
$[$\ion{N}{II}$]~\lambda6584$ &   $1.18 \pm 0.46$ \\ 
H$\alpha~\lambda6563$ & $8.35 \pm 0.57$ \\
$[$\ion{O}{III}$]~\lambda5007$ &  $3.71 \pm 0.68$ \\
$[$\ion{O}{III}$]~\lambda4959$ &  $1.22 \pm 0.46$ \\
H$\beta~\lambda4861$ & $2.25 \pm 0.58$ \\
$[$\ion{O}{II}$]~\lambda3729$ &  $ 5.65\pm 0.93$  \\
$[$\ion{O}{II}$]~\lambda3727$ &  $ 2.63 \pm 0.63$ \\ \hline
\end{tabular}
\end{table}

We measured the host galaxy extinction using the Balmer decrement, finding a value of ${\rm H}\alpha / {\rm H}\beta = 3.7 \pm 1.0$. This is larger than the theoretical ratio of 2.87 (assuming Case B recombination and a temperature of 10,000~K; \citealt{Osterbrock2006}); however, the noisy H$\beta$ measurement means it is also consistent with the theoretical value within the error. Assuming a \citet{Calzetti2000a} reddening law with $R_V = 3.1$, we estimate $E(B-V)_{\rm host} = 0.22^{+0.20}_{-0.22}~{\rm mag}$. This is consistent with the value from the SED modeling. Given the large uncertainty in the extinction and the dwarf nature of the host galaxy, we did not apply any host galaxy extinction correction to the SN photometry.

No auroral lines are detected in the spectrum of SN\,2020zbf, so we are limited to strong-line metallicity indicators. Given the large uncertainty in the extinction, we prefer indicators that use ratios of nearby lines and are therefore less sensitive to the host extinction, such as N2 or O3N2 \citep{Pettini2004}. Using the calibration of \citet{Marino2013} as implemented in the \program{pyMCZ} package \citep{Bianco2016}, we calculated $12 + \log({\rm O/H}) = 8.31^{+0.04}_{-0.05}$~dex using O3N2. Taking the solar value to be $12+\log({\rm O/H})_{\odot} = 8.69$~dex \citep{Asplund2021}, this corresponds to a metallicity $Z = 0.4~Z_{\odot}$.

We converted the H$\alpha$ flux to a star formation rate following ${\rm SFR} (\rm M_{\odot}~{\rm yr}^{-1})= 7.9 \times 10^{-42} \times$ $L\left(\rm H\alpha\right) {\rm erg~s}^{-1}$ \citep{Kennicutt1998}. This gives a star formation rate of 0.12~$\rm M_{\odot}~{\rm yr}^{-1}$ after correcting for extinction, and $0.07~\rm M_{\odot}~{\rm yr}^{-1}$ if assuming zero host extinction. This is slightly lower than, but consistent with, what is inferred from the SED modeling. 

Taken together, the properties of the host galaxy of SN\,2020zbf are quite typical of SLSN-I host galaxies (e.g., \citealt{Lunnan2014,Leloudas2015,Angus2016,Perley2016,Chen2017b,Schulze2018}). The luminosity, mass, and metallicity are all consistent with a dwarf galaxy and are near the center of the distribution seen for the hosts of these transients. Thus, even though some of the SN properties are unusual, the host galaxy environment is not.

\section{Discussion} \label{sec:discussion} 

\subsection{Powering mechanisms}

\subsubsection{Can SN\,2020zbf be powered by a magnetar?}

In Sects.~\ref{sec:synthetic_spectra} and~\ref{sec:lc_model}, comparisons of the X-shooter spectrum with tabulated spectra from physical SLSN models, as well as photometric modeling with MOSFiT, result in a preference toward a magnetar-powered model with a low ejecta mass (3.63~$\rm M_\odot$ and 1.51~$\rm M_\odot$, respectively).  We note that the synthetic spectra of \cite{Dessart2019} have not been modeled to match SN\,2020zbf or any other SLSN, but qualitatively match SN\,2020zbf well, which strengthens the hypothesis of a low ejecta-mass magnetar-powered event for SN\,2020zbf. The similarity with the magnetar properties estimated for typical SLSNe-I in \cite{Chen2023b} shows that the existence of low ejecta mass SLSNe-I powered by a magnetar is feasible. The correlation between the low ejecta mass and the high spin period ($P$ = 5.6~ms) in SN\,2020zbf is in agreement with the studies of \cite{Blanchard2020}, \cite{Hsu2021} and \cite{Chen2023b}, and indicates that low-mass ejecta SLSNe-I require less central power with slower-spinning neutron stars.

SN\,2020zbf presents unusually strong \ion{C}{II} emission lines in its early spectra, which likely require the energy from the magnetar to be absorbed in the outer layers where the C is more abundant. The low-mass ejecta could be a critical factor in producing these lines, for which the extra energy from the magnetar is not completely absorbed in the inner O-rich region, diffuses out and thermally excites the \ion{C}{}.

If the ejecta mass was the only key for the presence of strong \ion{C}{II} lines in the early spectra, we would expect all the low ejecta mass SLSNe-I to present strong \ion{C}{II} features, and these features to be absent for higher ejecta masses. In \cite{Dessart2019}, some of the models with higher ejecta masses ($M_{\rm ej}$ = 9.6~$\rm M_\odot$) and similar magnetar properties as SN\,2020zbf still present strong \ion{C}{II} lines, which \cite{Dessart2019} explains with the presence of a C-rich shell in the outermost layers of the ejecta. These lines also vanish as the photosphere recedes into the inner layers where C is less abundant. In addition, the three objects in the C-rich sample (SN\,2018bsz, SN\,2020wnt and SN\,2017gci) that have been fit with a magnetar model, cover a wide range of magnetar properties ($B$ = 2 -- 6~$\times~10^{14}$~G and $P$ = 2.8 -- 7~ms) and show a preference for higher ejecta masses ($M_{\rm ej}$ = 9 -- 26~M$_\odot$) than SN\,2020zbf. Thus, we speculate that different scenarios including the ejecta mass, the magnetar properties and, the mass fraction of C in the progenitor might be the keys in producing strong \ion{C}{II} lines in the spectra.

The main discrepancy in the \ion{C}{II} lines between the magnetar model and our observations is the lack of P-Cygni profiles seen in our spectra. However, based on the similarities of both the light curves and spectra, we conclude that the magnetar-powered explosion of a low-mass, C-rich progenitor star is a viable explanation for the observed properties of SN\,2020zbf.

\subsubsection{Can SN\,2020zbf be powered by CSM interaction?} \label{sec:csm_discussion}

Another mechanism that has been suggested to power SLSN-I light curves, is CSM interaction \citep[e.g.,][]{Chatzopoulos2012,Sorokina2016,Wheeler2017}, in which the kinetic energy of the ejecta is converted into radiation \citep[e.g.,][]{Zeldovich1967}. The exploration of this scenario is done under two assumptions; the \ion{C}{II} lines are pure emission and the energy deposition in the outer layers of the ejecta results from interaction rather than from a magnetar.

The absence of H and \ion{He}{} and the presence of strong \ion{C}{II} emission features in the spectra indicates a possible interaction of the SN ejecta with CO CSM 
\citep{Woosley2007,Blinnikov2010,Chevalier2011,Chatzopoulos2012,Quataert2012,Sorokina2016}. \cite{Chen2023b} estimated between 25\% and 44\% of SLSNe-I are preferentially fit by the H-poor CSM models. \cite{Sorokina2016}, using hydro-simulations, succeeded to reproduce the light curves of SN\,2010gx \citep{Pastorello2010} and PTF09cnd \citep{Quimby2011} with CSM interaction based on the ejecta mass, density structure, explosion energy, expansion of the CSM and C/O ratio. Moreover, \citet{Tolstov2017}, using radiation-hydrodynamics calculations, modeled PTF12dam and found that interaction with CSM ejected due to the pulsational pair-instability mechanism \citep{Barkat1967} can describe the light curve. 

\cite{Suzuki2020} found a systematic relation between peak bolometric luminosity and rise time depending on the kinetic energy, CSM mass, CSM radius and ejecta mass. SN\,2020zbf, if powered by interaction, falls in the regime of 1~$\rm M_\odot$ ejecta, 3~$\rm M_\odot$ CSM and 0.5 $\times$ 10$^{51}$ erg kinetic energy when considering its peak bolometric luminosity and the rise time. \cite{Khatami2023} defines four light-curve classes based on the parameters of the CSM/ejecta configuration and a qualitative comparison of SN\,2020zbf shows that it belongs to the interior breakout--heavy CSM  ($M_{\rm CSM}$ > $M_{\rm ej}$) class in which the shock breakout occurs within the CSM \citep{Ginzburg2012,Dessart2015}. In addition, the peak light -- rise time relation in \cite{Khatami2023} shows that for SN\,2020zbf the CSM mass is estimated between 1 and 10~$\rm M_\odot$ toward the lower limit.

\cite{Janka2012} showed that neutrino-driven SN explosions cannot explain kinetic energies larger than $\sim$ 2~$\times$~10$^{51}$~erg. Since the kinetic energy of SN\,2020zbf is estimated to be $\sim$ 0.8~$\times$~10$^{51}$~erg and assuming that the explosion mechanism in an interaction-dominated SN is driven by neutrinos, it is possible the main power source of SN\,2020zbf could be CSM interaction.

On the other hand, there are no spectral models that can fit the line shapes in the SLSN-I spectra in the case of interaction with CO CSM. The \ion{C}{II} line profiles in the spectra of SN\,2020zbf are asymmetric, showing a conspicuous tail extending to the red, which indicates a multiple electron scattering effect  \citep[e.g.,][]{Jerkstrand2017}. However, it is unknown what CSM configuration could give rise to the sharp peak in the \ion{C}{II} lines.

As discussed above, the \ion{C}{II} lines in SN\,2020zbf vanish with time except for the line at 6580~\AA. \cite{Pursiainen2022} studied the C-rich SLSN-I SN\,2018bsz and concluded that at $+24$ days post-maximum the \ion{C}{II} at 6580~\AA\ is replaced by H$\alpha$. Late hydrogen emission has also been observed in the C-rich SLSN-I PTF10aacg \citep{Yan2015} at $+75$ days post-maximum. Both objects have been explained with aspherical CSM interaction. Late-time interaction with spherical CSM resulting in H$\alpha$ emission has been seen in other SLSNe-I \citep[e.g][]{Yan2015,Yan2017,Fiore2021} and \cite{Yan2015} estimate that at least 15\% of SLSNe-I interact with previously ejected H-rich material at late times.

The evolution of the \ion{C}{II} lines of SN\,2020zbf shows a similar pattern as for SN\,2018bsz, but there is no obvious absorption component blueward of 6580~\AA\ in SN\,2020zbf. Furthermore, there is no indication of H$\beta$ and H$\gamma$ appearing,  but this can be limited by the low S/N of the spectra. However, due to the weakness of the other \ion{C}{II} lines, we can assume that this feature is more likely dominated by another ion rather than \ion{C}{II}. Additionally, the epoch of the studied spectrum falls in the period of a short ($\sim$ 15 days) plateau in the LCO light curve between $+30$ and $+45$ days past peak, which, if real, could be an indication of interaction. On the other hand, the shape of the SN\,2020zbf light curve could imply a combination of different mechanisms such as interaction early on followed by a magnetar power source. However, due to the possible presence of H$\alpha$ in the spectra, the plateau is more likely to be associated with interaction with H-rich material located at $\lesssim$ 4.5~$\times$~10$^{15}$~cm ($\rm v_{ej} \times 43.7 $~days; due to the low coverage this value is set as an upper limit). The multiple CSM layers are consistent with the pulsational pair instability model \citep{Woosley2017} that has been suggested as a mechanism for SLSNe \citep{Woosley2007}.

In reality, SN\,2020zbf could be powered by both a magnetar and CSM interaction. The preferred magnetar 5p11Bx2 model could well be correct giving a very good match with SN\,2020zbf spectra with a preference to low ejecta mass, but there could also be some CO CSM contributing mainly to the line profiles (sharply peaked emission \ion{C}{II} lines) and slightly to the light curve.

One possible way to break the degeneracy between the two powering mechanisms would be with radio observations. Both CSM interaction and magnetars are expected to produce detectable radio emission, but the timescales should be very different. Radio emission from CSM interaction should be produced within the first few years, while with a magnetar engine, the emission should not be detectable until around a decade \citep{Omand2018}.  Two SLSNe have been detected in radio so far, PTF10hgi \citep{Eftekhari2019, Law2019, Mondal2020, Hatsukade2021}, which is consistent with the magnetar model, and SN\,2017ens \citep{Margutti2023}, which is consistent with CSM interaction.  \cite{Omand2018} also shows that for the magnetar model, SLSNe with lower ejecta masses, similar to SN\,2020zbf, produce brighter radio emission at earlier times, which may make SN\,2020zbf a good candidate for follow-up observations.

\subsection{Progenitor}

Previous studies in the literature have suggested that the host galaxies of SLSNe-I have low metallicity and high star formation rate \citep[e.g.,][]{Leloudas2015} while \cite{Chen2017b} found that host galaxies of SLSN-I progenitors have a metallicity cut-off at 0.5 solar metallicity (see also \citealt{Perley2016,Schulze2018}).

Given the spectroscopic similarities of SLSNe-I with stripped-envelope SNe (\citealt{Pastorello2010}), there are two main scenarios for the progenitor stars; single WR stars \citep[e.g.,][]{Georgy2009} and stars in binary systems \citep[e.g.,][]{Nomoto1990,Yoon2010}. The possibly identified progenitors of SN-Ib iPTF13bvn \citep{Cao2013} and SN-Ic SN\,2017ein \citep{VanDyk2018} have been discussed in the context of both these scenarios \citep{Groh2013,Bersten2014,VanDyk2018}. In addition, \cite{Pursiainen2022} argue that for SN\,2018bsz both a single WR star and a star in a binary system are possible. \cite{Nicholl2017} show that SLSNe-I result from CO cores with $M_{\rm CO}$ $\geq$ 4~$\rm M_\odot$, which corresponds to $M_{\rm ZAMS}$ $\geq$ 20~$\rm M_\odot$ \citep{Yoon2010} comparing with the simulations of \cite{Yoon2006} at comparable metallicities. Rapid rotation might be the key to magnetar formation since the angular momentum needed to form a millisecond magnetar requires CO cores with initial rotational velocities > 200 -- 300~km~s$^{-1}$ \citep{Yoon2006}. On the other hand, \cite{DeMink2013} showed that binarity could also supply the necessary angular momentum to the stars, either through merging or Roche lobe overflow. \cite{Nicholl2017} argue that the rapid rotation of SLSN-I progenitors plays a crucial role, with the progenitor mass and the low metallicity being consequences of this.

SN\,2020zbf has been characterized as type Ic due to the absence of H and \ion{He}{} in the spectra, which is associated with H and possibly \ion{He}{} envelope losses in the progenitor star by stellar winds or mass transfer in a binary system before the explosion. In the case of a magnetar-powered event, the dominance of  \ion{C}{}  instead of  \ion{O}{}  in the spectra would potentially imply an initially high fraction of  \ion{C}{}  in the progenitor star. The best-fit 5p11Bx2 model (see Sect.\ref{sec:synthetic_spectra}) from \cite{Dessart2019} corresponds to an initial massive 60~$\rm M_\odot$ ZAMS star in a binary system that has stripped the hydrogen envelope through mass transfer during core hydrogen burning. It has lost the \ion{He}{} envelope, and its surface is within the C-rich part of the CO core \citep{Yoon2010,Dessart2015}. This progenitor could be associated with a carbon-type WC star that ends with a final mass of 5.11~$\rm M_\odot$ and it can create a SN-Ic with 3.63~$\rm M_\odot$ ejecta from which 0.89~$\rm M_\odot$ is C and 1.40~$\rm M_\odot$ is O. The appearance of strong \ion{C}{II} lines in the spectra of SN\,2020zbf around peak and their vanishing in the late-time spectra is consistent with the stratification of the 5p11 model, in which the outer ejecta are enriched in C. The modeling of the light curve with MOSFiT results in a progenitor pre-SN star of 3.13~$\rm M_\odot$ ($M_{\rm ej}$ + $M_{\rm NS}$). Both the light curve modeling and the spectral comparison show a low-mass progenitor star at explosion with masses between 3 and 5~$\rm M_\odot$, lower than the median value of 6.83$^{+4.04}_{-2.45}$~$\rm M_\odot$ in \cite{Chen2023b} but not unprecedented since there are a few progenitor systems in the ZTF sample with similar values to SN\,2020zbf.

In the case of the CSM interaction, the early photospheric spectra could point toward an interaction with CO material expelled before the explosion. The comparison with different CSM models shows that the progenitor star ($M_{\rm ej}$+ $M_{\rm CSM}$ + 1.4~$\rm M_\odot$, a typical value for a neutron star \citealt{Lattimer2007}) is between 6 and 10~$\rm M_\odot$ (6.4~$\rm M_\odot$; \citealt{Chatzopoulos2013}, 5.4~$\rm M_\odot$; \citealt{Suzuki2020}, < 10~$\rm M_\odot$; \citealt{Khatami2023}), lower than the value of 17.92$^{+24.11}_{-9.82}$~$\rm M_\odot$ in \cite{Chen2023b} but still not extraordinary. However, taking into account the preference of all the models in low ejecta mass (0.22 -- 3~$\rm M_\odot$), the fast rise time in the SN\,2020zbf's light curve, and the fact that the CSM might have been expelled well in advance and should not be considered in the mass of the pre-SN progenitor, we conclude that the progenitor star at explosion is estimated to be in the range of 2 -- 5~$\rm M_\odot$.

\cite{Anderson2018} studied the SLSN-I SN\,2018bsz in particular for the strong \ion{C}{II} and they concluded that these lines are produced by a magnetar-powered explosion of a massive C-rich WR progenitor (11.4~$\rm M_\odot$ pre-SN mass) in a binary system by comparing qualitatively with magnetar models from \cite{Dessart2019}. Considering the good match of the 5p11Bx2 model with the observed spectrum of SN\,2020zbf, a possible progenitor system for SN\,2020zbf could be as well a WR in a binary system. Binary interaction could also remove the H-envelope and form a H-rich CSM around SN\,2020zbf, with which the ejecta will interact at later times.

\subsection{Implications of C-rich SLSNe-I}

In this work, we have compared SN\,2020zbf with H-poor SLSNe that show \ion{C}{II} in emission. The comparison is not with a statistically complete sample of all C-dominated objects but rather with a careful selection of published C-rich objects. \cite{Anderson2018} suggest that the \ion{C}{II} lines in SN\,2018bsz stem from the large fraction of  \ion{C}{}  in the progenitor model. Even though we reach the same conclusion for SN 2020zbf, our SN presents stronger lines of \ion{C}{II}, possibly resulting from the lower ejecta mass. On the other hand, PTF10aagc seems to be the best spectral match for SN\,2020zbf, but no study for the powering mechanism of that SN has been done so far.

\cite{Fiore2021} fit SN\,2017gci photometry to synthetic light curves and model the nebular spectra and found that SN\,2017gci can be fit well by both a magnetar model and CSM interaction with an ejecta mass of 10~$\rm M_\odot$, although they do not comment on the progenitor's composition. In SN\,2020wnt, the high ejecta mass ($\sim$ 26~$\rm M_\odot$; \citealt{Tinyanont2023}) hides the magnetar at early times and it was not revealed until the nebular phase. The progenitor star for that SN was suggested to be a massive rotating star with high abundances of C in its outer layers \citep[][their fig. A2]{Tinyanont2023}.

The difference in the properties of the SLSN powering mechanisms reflects the large diversity in the light curves and the spectra of these objects. Carbon is always present in the outermost CO layers of the ejecta, and thus, we expect to see it in the spectra of SLSNe-I. However, the presence of lines stronger than for the typical SLSNe-I could imply a distinct mechanism and/or a specific composition in the ejecta. In any case, the presence of strong \ion{C}{II} could indicate a higher amount of C in the ejecta. A detailed analysis of a statistically meaningful C-rich SLSN-I sample is required in the future to understand the mechanisms for the formation of \ion{C}{II} lines and the properties of their progenitor systems.

\section{Conclusions} \label{sec:conclusions}

In this work we have studied the H-poor SLSN SN\,2020zbf, analyzing photometric data from ATLAS, LCO, and UVOT and spectra taken with NTT+EFOSC and
the VLT+X-shooter instrument. Our main conclusions are as follows:
\begin{itemize}
    \item SN\,2020zbf is a fast-rising SLSN-Ic with a rise time of $\lesssim$ 26.4 days from the explosion and a peak magnitude of $M_{\rm g} = -21.2$~mag. The rise time is on the faster end of the distribution for SLSNe-I.
    \item The early spectra of SN\,2020zbf present three strong \ion{C}{II} lines that are not typically seen in normal SLSNe-I.
    \item Both the light curve modeling and the comparison with synthetic spectra are consistent with a magnetar-powered SN of a C-rich star with an ejecta mass of about 1.5 -- 3~$\rm M_\odot$.
    \item Alternatively and/or additionally, we argue that the shape and the strength of the \ion{C}{II} lines can also be attributed to an interaction between low-mass ejecta and a dense CO CSM.
    \item Based on the above modeling, the progenitor is estimated to have a pre-SN mass of between 2 and 5~$\rm M_\odot$.
    \item A potential late-time H$\alpha$ emission that is accompanied by a knee in the LCO light curve could be an indication of interaction with H-rich CSM. 
    \item The host galaxy has a mass of log(M/$\rm M_\odot$) = $8.68^{+0.18}_{-0.22}$, a star formation rate of $0.24^{+0.41}_{-0.12}~\rm M_\odot\,{\rm yr}^{-1}$, and a metallicity of 0.4~$\rm Z_\odot$, similar to typical SLSN-I host galaxies.
    \item There is large variety in the light curves and the spectra of SLSNe with strong \ion{C}{II}, suggesting that many different configurations can result in such a spectral signature.
\end{itemize}

\noindent This object illustrates the challenges in classifying SLSNe-I by demonstrating how diverse even the SLSN class can be, as well as the implications for both progenitor populations and explosion mechanisms. More sophisticated tools for light curve and spectra modeling are required to explain the peculiarities of similar objects.

\begin{acknowledgements}

The authors would like to thank the anonymous referee for their comments and suggestions. Based on observations collected at the European Organisation for Astronomical Research in the Southern Hemisphere, Chile, as part of ePESSTO+ (the advanced Public ESO Spectroscopic Survey for Transient Objects Survey).
ePESSTO+ observations were obtained under ESO program IDs 106.216 and 108.220C (PI: Inserra).
LCO data have been obtained via an OPTCON proposal (IDs: OPTICON 20A/015, 20B/003, 21B/001, 22B/ 002) and as part of the Global Supernova Project. The OPTICON project has received funding from the European Union’s Horizon 2020 research and innovation programme under grant agreement No 730890.
This work makes use of data from the Las Cumbres Observatory global network of telescopes.  The LCO group is supported by NSF grants AST-1911151 and AST-1911225.
This work has made use of data from the Asteroid Terrestrial-impact Last Alert System (ATLAS) project. ATLAS is primarily funded to search for near earth asteroids through NASA grants NN12AR55G, 80NSSC18K0284, and 80NSSC18K1575; byproducts of the NEO search include images and catalogs from the survey area. The ATLAS science products have been made possible through the contributions of the University of Hawaii Institute for Astronomy, the Queen’s University Belfast, the Space Telescope Science Institute, and the South African Astronomical Observatory. 
R. Lunnan is supported by the European Research Council (ERC) under the European Union’s Horizon Europe research and innovation programme (grant agreement No. 10104229 - TransPIre).
S. Schulze acknowledges support from the G.R.E.A.T. research environment, funded by {\em Vetenskapsr\aa det},  the Swedish Research Council, project number 2016-06012. 
M. Nicholl is supported by the European Research Council (ERC) under the European Union’s Horizon 2020 research and innovation programme (grant agreement No.~948381) and by UK Space Agency Grant No.~ST/Y000692/1. 
T.-W. Chen acknowledges the Yushan Young Fellow Program by the Ministry of Education, Taiwan for the financial support. 
M. Pursiainen acknowledges support from a UK Research and Innovation Fellowship (MR/T020784/1)
T. E. M\"uller-Bravo acknowledges financial support from the  
Spanish Ministerio de Ciencia e Innovaci\'on (MCIN), the Agencia  
Estatal de Investigaci\'on (AEI) 10.13039/501100011033, and the  
European Union Next Generation EU/PRTR funds under the 2021 Juan de la  
Cierva program FJC2021-047124-I and the PID2020-115253GA-I00 HOSTFLOWS  
project, from Centro Superior de Investigaciones Cient\'ificas (CSIC)  
under the PIE project 20215AT016, and the program Unidad de Excelencia  
Mar\'ia de Maeztu CEX2020-001058-M.
This work was funded by ANID, Millennium Science Initiative, ICN12\_009.
G. Pignata acknowledge support from ANID through Millennium Science Initiative Programs ICN12\_009.
C. P. Guti\'errez acknowledges financial support from the Secretary of Universities
and Research (Government of Catalonia) and by the Horizon 2020 Research
and Innovation Programme of the European Union under the Marie
Sk\l{}odowska-Curie and the Beatriu de Pin\'os 2021 BP 00168 programme,
from the Spanish Ministerio de Ciencia e Innovaci\'on (MCIN) and the
Agencia Estatal de Investigaci\'on (AEI) 10.13039/501100011033 under the
PID2020-115253GA-I00 HOSTFLOWS project, and the program Unidad de
Excelencia Mar\'ia de Maeztu CEX2020-001058-M.

\end{acknowledgements}

\bibliographystyle{aa}
\bibliography{references}

\begin{appendix}

\section{Photometric data} \label{app:photometry}

\begin{table}[h]
    \centering
    \caption{Photometric data of SN\,2020zbf.}
    
    \label{tab:phot_data}
    \small
    \begin{tabular}{ccccc}
    \hline
    \hline
     MJD & Phase\tablefootmark{\scriptsize a} & Filter & Magnitude & Instrument\tablefootmark{\scriptsize b}\\
     & (days) & & (mag) & \\
    \hline
59137.5 & -22.60 & $c$ & $20.89 \pm 0.35$ & ATLAS\\
59145.4 & -15.90 & $c$ & $19.65 \pm 0.10$ & ATLAS\\
59147.3 & -14.23 & $o$ & $19.73 \pm 0.21$ & ATLAS\\
59157.4 & -5.86 & $o$ & $18.96 \pm 0.18$ & ATLAS\\
59161.4 & -2.51 & $o$ & $18.87 \pm 0.08$ & ATLAS\\
59163.4 & -0.84 & $o$ & $18.93 \pm 0.07$ & ATLAS\\
59164.8 & 0 & $B$ & $18.64 \pm 0.20$ & LCO\\
59164.8 & 0 & $g$ & $18.60 \pm 0.07$ & LCO\\
59164.8 & 0 & $V$ & $18.66 \pm 0.06$ & LCO\\
59164.8 & 0 & $r$ & $18.75 \pm 0.10$ & LCO\\
59164.8 & 0 & $i$ & $18.90 \pm 0.07$ & LCO\\
59169.3 & 4.19 & $c$ & $18.79 \pm 0.05$ & ATLAS\\
59170.8 & 5.02 & $B$ & $18.75 \pm 0.14$ & LCO\\
59170.8 & 5.02 & $g$ & $18.76 \pm 0.09$ & LCO\\
59170.8 & 5.02 & $V$ & $18.85 \pm 0.14$ & LCO\\
59170.8 & 5.02 & $r$ & $18.97 \pm 0.04$ & LCO\\
59170.8 & 5.02 & $i$ & $19.28 \pm 0.13$ & LCO\\
59176.1 & 10.04 & $B$ & $18.87 \pm 0.13$ & LCO\\
59176.1 & 10.04 & $g$ & $18.99 \pm 0.06$ & LCO\\
59176.1 & 10.04 & $V$ & $18.98 \pm 0.11$ & LCO\\
59176.1 & 10.04 & $r$ & $19.04 \pm 0.08$ & LCO\\
59176.1 & 10.04 & $i$ & $19.14 \pm 0.08$ & LCO\\
59178.01 & 11.73 & $UVW1$ & $20.50 \pm 0.15$ & \textit{Swift}/UVOT\\
59178.01 & 11.73 & $U$ & $19.04 \pm 0.11$ & \textit{Swift}/UVOT\\
59178.01 & 11.73 & $B$ & $18.99 \pm 0.14$ & \textit{Swift}/UVOT\\
59178.01 & 11.73 &      $UVW2$ &        $21.67 \pm 0.20$ & \textit{Swift}/UVOT\\
59178.01 & 11.73 &      $V$ &   $18.90 \pm 0.24$ & \textit{Swift}/UVOT\\
59178.01 & 11.73 &      $UVM2$ &        $21.27 \pm 0.15$ & \textit{Swift}/UVOT\\
59179.66 & 13.11 &      $UVW1$ &        $20.62 \pm 0.16$ & \textit{Swift}/UVOT\\
59179.66 & 13.11 &      $U$ &   $19.14 \pm 0.11$ & \textit{Swift}/UVOT\\
59179.66 & 13.11 &      $B$ &   $19.12 \pm 0.15$ & \textit{Swift}/UVOT\\
59179.67 & 13.11 &      $UVW2$ &        $21.33 \pm 0.17$ & \textit{Swift}/UVOT\\
59179.67 & 13.11 & $V$ &        $18.81 \pm 0.22$ & \textit{Swift}/UVOT\\
59179.67 & 13.11 &      $UVM2$ &        $21.36 \pm 0.16$ & \textit{Swift}/UVOT\\
59181.0 & 14.23 & $o$ & $18.89 \pm 0.17$ & ATLAS\\
59183.0 & 15.90 & $o$ & $19.09 \pm 0.22$ & ATLAS\\
59185.0 & 17.58 & $o$ & $19.16 \pm 0.16$ & ATLAS\\
59185.38 & 17.90 &      $UVW1$ &        $21.20 \pm 0.24$ & \textit{Swift}/UVOT\\
59185.38 & 17.90 &      $U$ &   $19.31 \pm 0.13$ & \textit{Swift}/UVOT\\
59185.38 & 17.90 &      $B$ &   $19.27 \pm 0.18$ & \textit{Swift}/UVOT\\
59185.38 & 17.90 &      $UVW2$ &        $22.62 \pm 0.38$ & \textit{Swift}/UVOT\\
59185.38 & 17.90 &      $V$ &   $18.66 \pm 0.22$ & \textit{Swift}/UVOT\\
59185.39 & 17.90 &      $UVM2$ &        $21.52 \pm 0.19$ & \textit{Swift}/UVOT\\
59186.1 & 18.41 & $B$ & $19.16 \pm 0.09$ & LCO\\
59186.1 & 18.41 & $g$ & $19.16 \pm 0.07$ & LCO\\
59186.1 & 18.41 & $V$ & $19.14 \pm 0.07$ & LCO\\
59186.1 & 18.41 & $r$ & $19.26 \pm 0.06$ & LCO\\
59186.1 & 18.41 & $i$ & $19.38 \pm 0.05$ & LCO\\
59187.3 & 19.25 & $o$ & $19.29 \pm 0.14$ & ATLAS\\
59191.3 & 22.60 & $o$ & $19.65 \pm 0.14$ & ATLAS\\
59193.9& 24.27 & $B$ & $19.28 \pm 0.11$ & LCO\\
59193.9& 24.27 & $g$ & $19.30 \pm 0.16$ & LCO\\
59193.9& 24.27 & $V$ & $19.30 \pm 0.13$ & LCO\\
59193.9& 24.27 & $r$ & $19.11 \pm 0.15$ & LCO\\
59193.9& 24.27 & $i$ &$ 19.34 \pm 0.19$ & LCO\\

\hline
\end{tabular}
\end{table}
\addtocounter{table}{-1}
\begin{table}
    \centering
    \captionsetup{list=no}
    \caption{continued.}
    \small
    \begin{tabular}{ccccc}
    \hline
    \hline
     MJD & Phase\tablefootmark{\scriptsize a} & Filter & Magnitude & Instrument\tablefootmark{\scriptsize b}\\
     & (days) & & (mag) & \\
    \hline
    59193.27 & 24.50 &  $UVW1$ &        $21.34 \pm 0.38$ & \textit{Swift}/UVOT\\
59193.27 & 24.50 &      $U$ &   $19.58 \pm 0.22$ & \textit{Swift}/UVOT\\
59193.27 & 24.50 &      $B$ &   $19.47 \pm 0.30$ & \textit{Swift}/UVOT\\
59193.28 & 24.50 &      $UVW2$ &        $22.46 \pm 0.53$ & \textit{Swift}/UVOT\\
59193.28 & 24.50 &      $V$ &   $18.73 \pm 0.3$4 & \textit{Swift}/UVOT\\
59193.28 & 24.50 &      $UVM2$ &        $21.85 \pm 0.33$ & \textit{Swift}/UVOT\\
59193.3 & 24.27 & $c$ & $19.27 \pm 0.10$ & ATLAS\\
59195.3 & 25.95 & $c$ & $19.50 \pm 0.10$ & ATLAS\\
59197.3 & 27.62 & $c$ & $19.43 \pm 0.12$ & ATLAS\\
59201.2 & 30.97 & $B$ & $19.57 \pm 0.08$ & LCO\\
59201.2 & 30.97 & $g$ & $19.47 \pm 0.04$ & LCO\\
59201.2 & 30.97 & $V$ & $19.35 \pm 0.07$ & LCO\\
59201.2 & 30.97 & $r$ & $19.39 \pm 0.08$ & LCO\\
59201.2 & 30.97 & $i$ & $19.66 \pm 0.07$ & LCO\\
59203.3 & 32.64 & $o$ & $19.45 \pm 0.09$ & ATLAS\\
59205.3 & 34.32 & $o$ & $19.59 \pm 0.17$ & ATLAS\\
59207.5 & 35.99 & $B$ & $19.65 \pm 0.19$ & LCO\\
59207.5 & 35.99 & $V$ & $19.34 \pm 0.22$ & LCO\\
59207.5 & 35.99 & $o$ & $19.41 \pm 0.12$ & ATLAS\\
59211.3 & 39.34 & $o$ & $19.73 \pm 0.27$ & ATLAS\\
59213.3 & 41.01 & $o$ & $19.49 \pm 0.28$ & ATLAS\\
59215.2 & 42.69 & $g$ & $19.60 \pm 0.14$ & LCO\\
59215.2 & 42.69 & $V$ & $19.49 \pm 0.17$ & LCO\\
59215.2 & 42.69 & $r$ & $19.48 \pm 0.12$ & LCO\\
59215.2 & 42.69 & $i$ & $19.42 \pm 0.12$ & LCO\\
59219.3 & 46.04 & $c$ & $19.52 \pm 0.09$ & ATLAS\\
59221.1 & 47.71 & $B$ & $19.83 \pm 0.15$ & LCO\\
59221.1 & 47.71 & $g$ & $19.65 \pm 0.28$ & LCO\\
59221.1 & 47.71 & $V$ & $19.52 \pm 0.09$ & LCO\\
59221.1 & 47.71 & $r$ & $19.53 \pm 0.20$ & LCO\\
59221.1 & 47.71 & $i$ & $19.60 \pm 0.10$ & LCO\\
59227.1 & 52.73 & $B$ & $19.98 \pm 0.14$ & LCO\\
59227.1 & 52.73 & $g$ & $19.87 \pm 0.10$ & LCO\\
59227.1 & 52.73 & $V$ & $19.66 \pm 0.13$ & LCO\\
59227.1 & 52.73 & $r$ & $19.65 \pm 0.15$ & LCO\\
59227.1 & 52.73 & $i$ & $19.74 \pm 0.21$ & LCO\\
59227.1 & 52.73 & $c$ & $19.87 \pm 0.12$ & ATLAS\\
59229.2 & 54.40 & $c$ & $19.75 \pm 0.13$ & ATLAS\\
59231.3 & 56.08 & $c$ & $19.43 \pm 0.10$ & ATLAS\\
59233.1 & 57.76 & $B$ & $20.12 \pm 0.15$ & LCO\\
59233.1 & 57.76 & $g$ & $20.01 \pm 0.11$ & LCO\\
59233.1 & 57.76 & $V$ & $19.79 \pm 0.13$ & LCO\\
59233.1 & 57.76 & $r$ & $19.77 \pm 0.19$ & LCO\\
59233.1 & 57.76  & $i$ & $19.85 \pm 0.25$ & LCO\\
59239.1 & 62.78 & $g$ & $20.18 \pm 0.21$ & LCO\\
59239.1 & 62.78 & $V$ & $19.96 \pm 0.22$ & LCO\\
59239.1 & 62.78 & $r$ & $19.90 \pm 0.18$ & LCO\\
59239.1 & 62.78 & $i$ & $20.10 \pm 0.12$ & LCO\\
59245.2 & 67.80 & $o$ & $20.07 \pm 0.26$ & ATLAS\\
59246.8 & 68.64 & $B$ & $20.54 \pm 0.28$ & LCO\\
59246.8 & 68.64 & $g$ & $20.38 \pm 0.15$ & LCO\\
59246.8 & 68.64 & $V$ & $20.19 \pm 0.22$ & LCO\\
59258.8 & 78.68 & $V$ & $20.39 \pm 0.28$ & LCO\\
59258.8 & 78.68  & $r$ & $20.29 \pm 0.21$ & LCO\\
59266.4 & 85.38 & $r$ & $20.36 \pm 0.20$ & LCO\\
59377.4 & 178.29 & $r$ & $21.62 \pm 0.23$ & LCO\\
59377.4 & 178.29 & $i$ & $21.79 \pm 0.26$ & LCO\\
\hline
   \end{tabular}
   \tablefoot{The photometry is reported on the AB system and is not corrected for reddening. This table is available in machine readable form. Multiple exposures on any given night are averaged to give the values presented here.\tablefoottext{a}{Rest-frame relative to the g-band maximum (MJD 59\,164.8).}\tablefoottext{b}{UVOT photometry is not host galaxy subtracted.}}
\end{table}

\newpage

\onecolumn
\section{Spectroscopic data} \label{app:spectra}

\begin{table*}[h]

\centering
\caption{SN\,2020zbf spectroscopic observations.}\label{tab:spectra}
\begin{tabular*}{.98\linewidth}[H]{@{\extracolsep{\fill}}ccccccc}
\hline
\hline
UT date & MJD & Phase\tablefootmark{\scriptsize a}  & Telescope + & Exposure & Grism & Wavelength range\\
 & (days) & & Instrument & (s) & & (\AA)\\
\hline
20201109 & 59162 & -2.37 & NTT + EFOSC2 & 900 & Gr\#13 & 3650 – 9250 \\
20201116 & 59168 & 2.65 & NTT  +EFOSC2 & 1800 & Gr\#11 + Gr\#16 & 3345 – 9995  \\
20201118 & 59170.8 & 4.33 & VLT + X-shooter & 2400 & -- & 3000 – 24800\\
20201208 & 59190 & 21.93 &      NTT + EFOSC2 &  2700 & Gr\#13 & 3650 – 9250\\
20201223 & 59205 & 33.62 &      NTT + EFOSC2 &  2700 & Gr\#13 & 3650 – 9250\\
20210104 & 59217 & 43.67 &      NTT + EFOSC2 &  2700 & Gr\#13 & 3650 – 9250\\
20210123 & 59233.1 & 57.06 &    NTT + EFOSC2 &  2700 & Gr\#13 & 3650 – 9250\\
\hline
\end{tabular*}
\tablefoot{\tablefoottext{a}{Rest-frame relative to the g-band maximum (MJD 59164.8).}}
\end{table*}

\FloatBarrier

\newpage

\section{MOSFiT results} \label{app:mosfit}

\begin{figure*}[h]
    \centering
    \includegraphics[width=1\columnwidth]{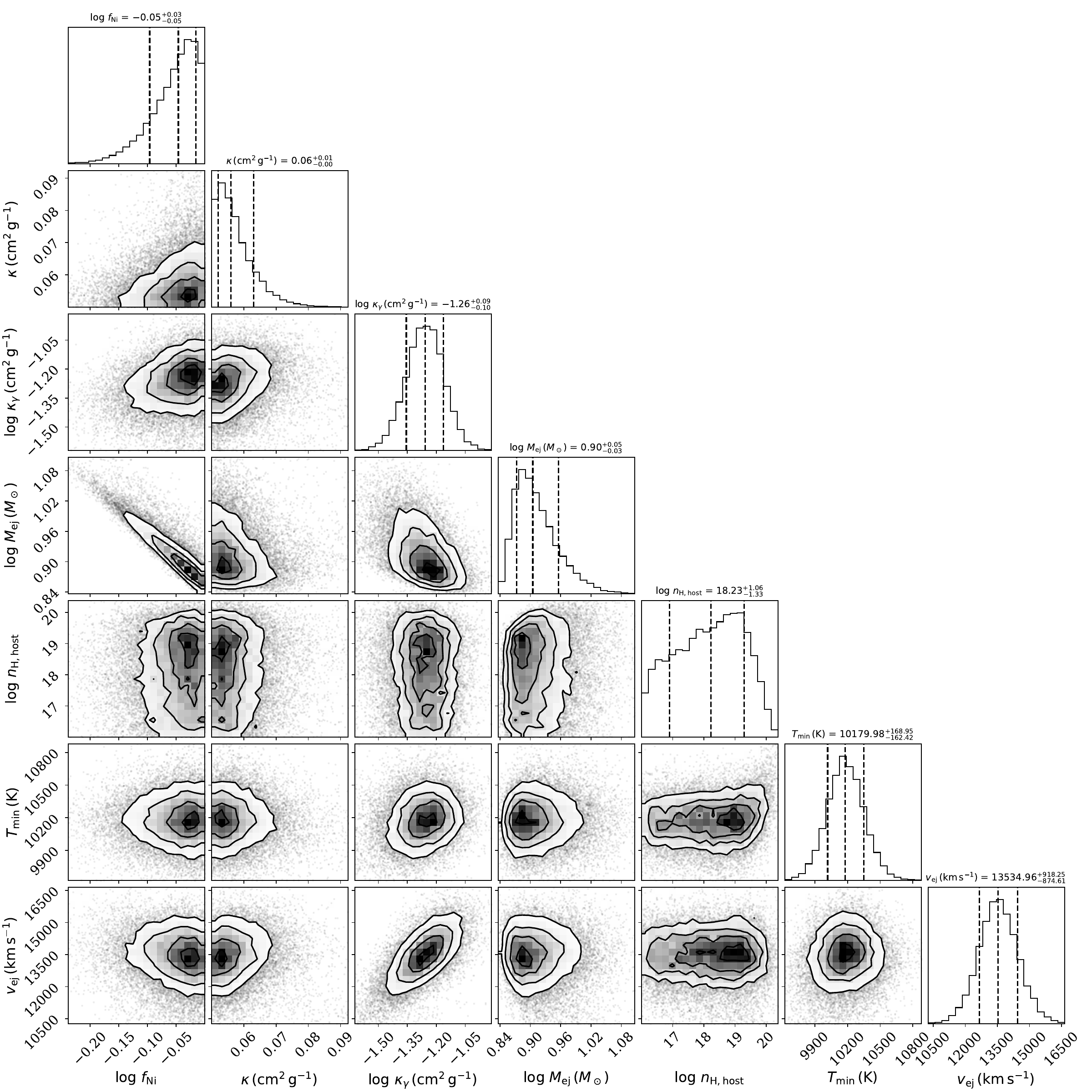}
    \caption{1D and 2D posterior distributions of the \program{default} $^{56}$Ni model parameters from the MOSFiT model. Median and 1$\sigma$ of the best fit values are marked and labeled.}
    \label{fig:corner_ni}
\end{figure*}

\begin{figure*}[h]
    \centering
    \includegraphics[width=1\columnwidth]{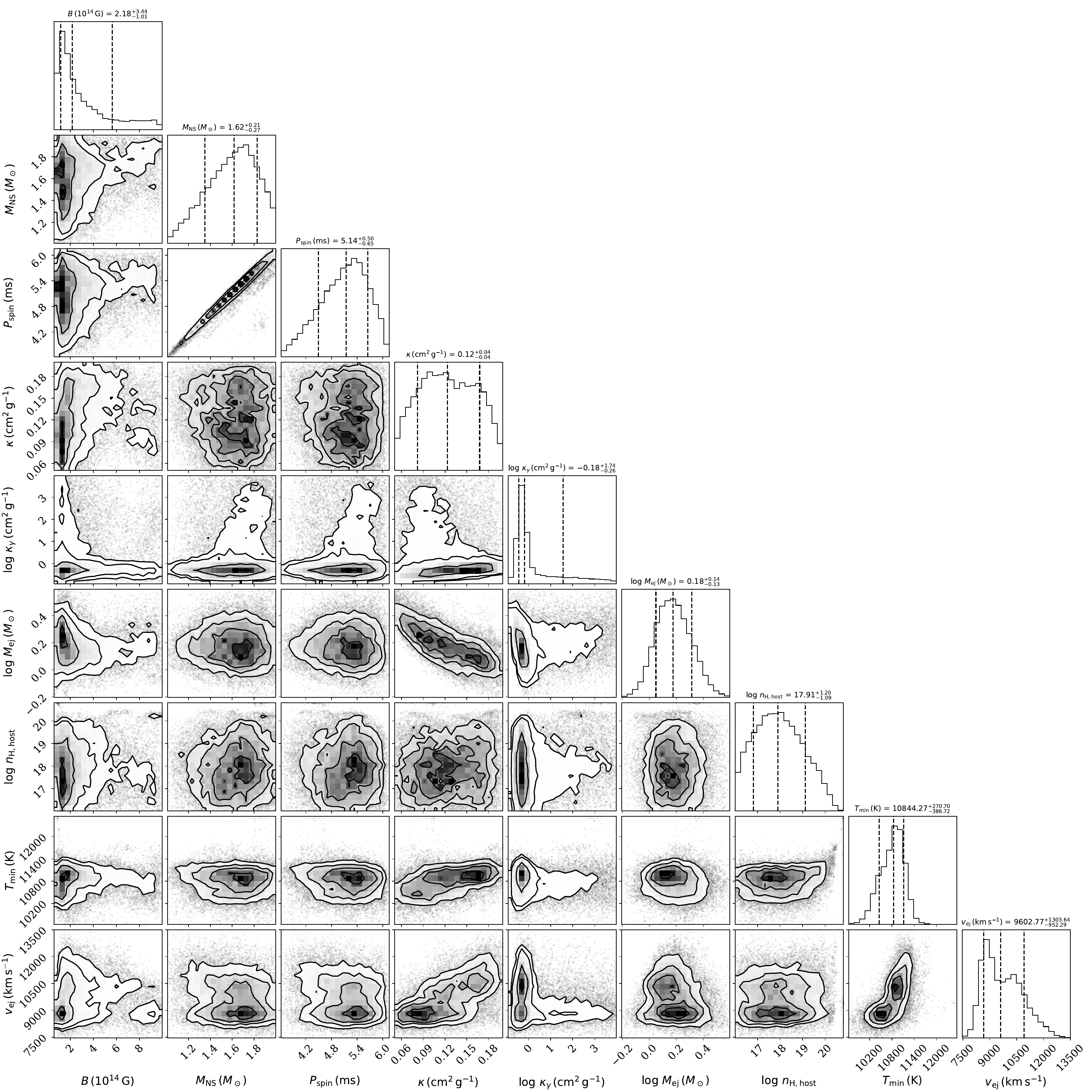}
    \caption{1D and 2D posterior distributions of the \program{slsn} magnetar model parameters from the MOSFiT model. Median and 1$\sigma$ of the best fit values are marked and labeled.}
    \label{fig:corner_slsn}
\end{figure*}

\begin{figure*}[h]
    \centering
    \includegraphics[width=1\columnwidth]{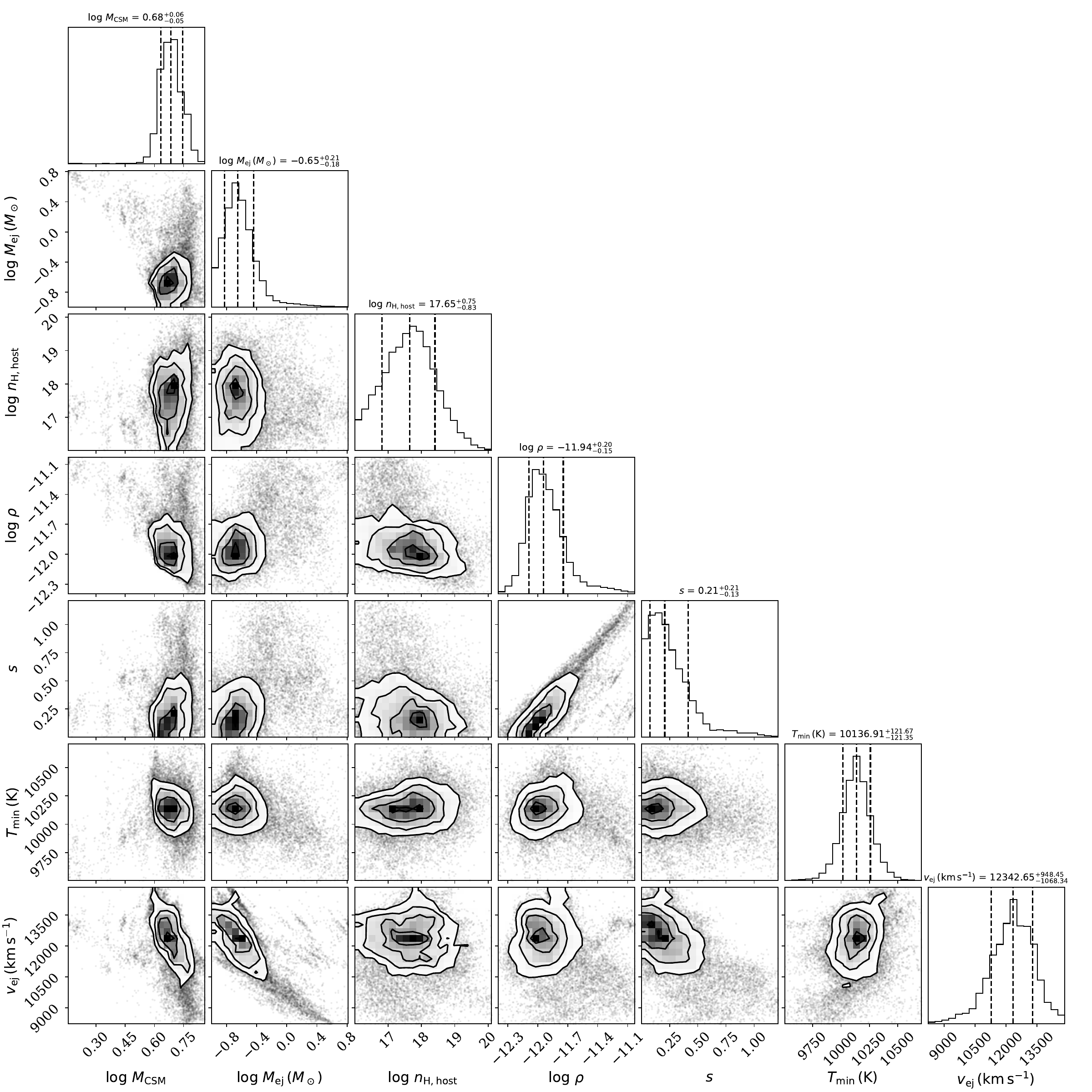}
    \caption{1D and 2D posterior distributions of the \program{csm} model parameters from the MOSFiT model. Median and 1$\sigma$ of the best fit values are marked and labeled.}
    \label{fig:corner_csm}
\end{figure*}

 \end{appendix}
\end{document}